\documentclass[titlepage,12pt]{article}
\usepackage{amsmath,amssymb,amscd,graphicx,epsfig}
\newcommand{\be}{\begin{equation}}
\newcommand{\ee}{\end{equation}}
\begin{document}
\addtolength{\baselineskip}{1mm}
\setlength{\parskip}{.5mm}

\addtolength{\abovedisplayskip}{1.5mm}
\addtolength{\belowdisplayskip}{1.5mm}

\begin{titlepage}
\begin{center}

{\hbox to\hsize{
\hfill PUPT-2209,~DUKE-CGTP-06-03}}

{\hbox to\hsize{\hfill hep-th/0610007}}

\vspace{3cm}

{\Large \bf D-branes at Singularities,\\[6mm]
 Compactification, 
and Hypercharge}\\[1.5cm]

{\large Matthew Buican${}^a$, Dmitry Malyshev${}^{a,}\footnote{On leave from ITEP Russia, Moscow, B Cheremushkinskaya, 25}$, David R. Morrison${}^{b,c}$,
 \\[4mm] Herman Verlinde${}^a$ and Martijn Wijnholt${}^{a,d}$}\\[8mm]

${}^a$ {\it Department of Physics, Princeton University, Princeton,
NJ 08544}\\[2mm]
${}^b$ {\it Center for Geometry and Theoretical Physics, Duke University, Durham,
NC 27708}\\[2mm]
${}^c$ {\it Departments of Physics and Mathematics, Univ.\ of California, 
Santa Barbara, CA 93106}\\[2mm]
${}^d$ {\it Max-Planck-Institut f\"ur Gravitationsphysik, 
Albert-Einstein-Institut, 
Potsdam, Germany}
\vspace*{2.2cm}

\end{center}
\noindent
We report on progress towards the construction of SM-like gauge theories on
the world-volume of D-branes at a Calabi--Yau singularity. In particular, we work
out the topological conditions on the embedding of the singularity inside a
compact CY threefold, that select hypercharge as the only light $U(1)$ gauge factor. 
We apply this insight to the proposed open string realization of the SM of hep-th/0508089, 
based on a D3-brane at a $dP_8$ singularity, and present a 
geometric construction of a compact Calabi--Yau threefold with all the required 
topological properties. We comment on the relevance of D-instantons to the breaking
of global $U(1)$ symmetries in D-brane models.

\end{titlepage}
\newpage
\tableofcontents

\renewcommand{\footnotesize}{\small}
\newpage

\newcommand{\ba}{\begin{eqnarray}
\addtolength{\abovedisplayskip}{1.2mm}
\addtolength{\belowdisplayskip}{1.2mm}}
\newcommand{\ea}{\end{eqnarray}}
\newcommand{\bea}{\begin{eqnarray*}}
\newcommand{\eea}{\end{eqnarray*}}

\newcommand{\FF}{{\mathsf F}}
\newcommand{\sss}{{\! s}}
\newcommand{\aaa}{{\mbox{\scriptsize \sc a}}}
\newcommand{\bbb}{{\mbox{\scriptsize \sc b}}}
\newcommand{\N}{{\mathcal{N}}}
\newcommand{\D}{{\mathcal{D}}}
\newcommand{\La}{{\mathcal{L}}}
\newcommand{\OO}{{\mathcal{O}}}

\newcommand{\ub}{\underbrace}

\newcommand{\newsection}[1]{
\addtocounter{section}{1} 
\setcounter{subsection}{0} \addcontentsline{toc}{section}{\protect
\numberline{\arabic{section}}{{\rm #1}}} \vglue .0cm \pagebreak[3]
\noindent{\large \bf  \thesection. #1}\nopagebreak[4]\par\vskip .3cm}
\newcommand{\newsubsection}[1]{
\addtocounter{subsection}{1}
\addcontentsline{toc}{subsection}{\protect
\numberline{\arabic{section}.\arabic{subsection}}{ #1}} \vglue .0cm
\pagebreak[3] \noindent{\it \thesubsection. #1}\nopagebreak[4]\par\vskip .3cm}

\newcommand{\Z}{{\mathbb{Z}}}
\newcommand{\R}{{\mathbb{R}}}
\newcommand{\C}{{\mathbb{C}}}
\newcommand{\PP}{{\mathbb{P}}}
\newcommand{\HH}{{\mathcal{H}}}
\newcommand{\Hd}{{\mathcal{H}}^*}
\newcommand{\lb}{\label}
\newcommand{\g}{\gamma}
\newcommand{\G}{\Gamma}
\newcommand{\ra}{\rightarrow}
\newcommand{\lra}{\longrightarrow}
\newcommand{\overar}{\overrightarrow}
\newcommand{\wt}{\widetilde}
\newcommand{\td}{\tilde}
\newcommand{\e}{\epsilon}
\newcommand{\al}{\alpha}
\newcommand{\bt}{\beta}
\newcommand{\p}{\partial}
\newcommand{\dl}{\delta}
\newcommand{\Dl}{\Delta}
\newcommand{\ld}{\lambda}
\newcommand{\Ld}{\Lambda}
\newcommand{\vp}{\varphi}
\newcommand{\te}{\theta}
\newcommand{\om}{\omega}
\newcommand{\sm}{\sigma}
\newcommand{\Sm}{\Sigma}
\newcommand{\ccap}{\cdot}

\vspace{-9mm}

\newsection{Introduction}

D-branes near Calabi--Yau singularities provide open string realizations of an increasingly rich class
of gauge theories \cite{ALE,KW,MoPl}.
Given the hierarchy between the Planck and TeV scale, it is natural to make use
of this technology and pursue a bottom-up 
approach to string phenomenology, that aims to find 
Standard Model-like theories on D-branes near CY singularities.
In this setting, the D-brane world-volume theory can be isolated from the closed string physics
in  the bulk via a formal decoupling limit, in which the string and 4-d Planck scale are taken to 
infinity, or very large.
The clear advantage of this bottom-up
strategy is that it separates the task of finding local realizations of SM-like models from the
more difficult challenge of finding fully consistent, realistic string compactifications.\footnote{
The general challenge of extending local brane constructions near CY singularities to 
full-fledged string compactifications represents a geometric component of the 
 ``swampland program'' of \cite{swamp},  that aims to determine the full class of quantum field 
theories that admit consistent UV completions with gravity.}

In scanning the space of CY singularities for candidates that 
lead to realistic gauge theories, one is aided by the fact that all gauge invariant 
couplings of the world-volume theory are controlled by the local geometry;
in particular, symmetry breaking patterns can be enforced by appropriately 
dialing the volumes of compact cycles of the singularity. 
Several other 
properties of the gauge theory, however, such as the spectrum of light $U(1)$ 
vector bosons and the number of freely tunable couplings, depend on 
how the local singularity is embedded inside the full compact 
Calabi--Yau geometry. 
 
 In this paper we work out some concrete aspects of this program.
We begin with a brief review of the general set of ingredients that can be used to build
semi-realistic gauge theories from branes at singularities. Typically these local
constructions lead to models
with extra $U(1)$ gauge symmetries beyond hypercharge. 
As our first new result, we identify the general 
topological conditions on the embedding of a CY singularity  inside a compact CY threefold,
that determines which 
$U(1)$-symmetry factors survive as massless gauge symmetries. 
The other $U(1)$ bosons acquire a mass of order of the string scale.
The left-over global symmetries are broken by D-brane instantons.

In the second half of the paper, we apply this insight to the concrete construction of an 
SM-like theory  given in \cite{MH},  based on a single D3-brane near
a suitably chosen del Pezzo 8 singularity. We specify a simple topological condition 
on the compact  embedding of the $dP_8$ singularity, such that only hypercharge survives as the massless gauge symmetry. To state this condition,
recall that the 2-homology of a $dP_8$ surface is spanned by
the canonical class $K$ and eight 2-cycles $\alpha_i$ with
intersection form $\alpha_i \ccap \alpha_j = - A_{ij}$
with $A_{ij}$ the Cartan matrix of $E_8$.  
With this notation, our geometric proposal
is summarized as follows:

\smallskip

\noindent
The world-volume gauge theory on a single D3-brane near a del Pezzo 8 singularity embedded in
a compact Calabi-Yau threefold with the following geometrical properties:\\[.3mm]
${}$ \ \  \ (i) the two 2-cycles $\alpha_1$ and $\alpha_2$ are degenerate and form a curve of $A_2$ singularities\\[.2mm]
${}$ \ \  \ (iii) all 2-cycles except $\alpha_4$ are non-trivial within the full Calabi-Yau three-fold\\[.2mm]
has, for a suitable choice of K\"ahler moduli, the gauge group and matter content of the SSM
\begin{table}[ht]\label{sm}
 \begin{center}
  {\renewcommand{\arraystretch}{1.1}
 \begin{tabular}{c c c c c c c c c c}
   \hline
     \rule[5mm]{0mm}{0pt} & $Q_i$ & $u_i^c$ & $d_i^c$ & $\ell_i$ & \ $e_i^c$ \ & \ \ $\nu_i^c$\ \ &\  $ H_i^u $ & $H_i^d$ \\
  \hline
\rule[5mm]{0mm}{0pt}
   $SU(3)_{C}$ \ \ &  \bf{3} & $\bar{\bf{3}}$ & $\bar{\bf{3}}$ &
    \bf{1} &  \bf{1} &  \bf{1} &  \bf{1}&   \bf{1} \\
    \rule[5mm]{0mm}{0pt}
  $ SU(2)_{L}$ \  &  \bf{2} &  \bf{1} &  \bf{1} &
    \bf{2} &  \bf{1} &  \bf{1}  &  \bf{2}  &   \bf{2}  \\
   \rule[5mm]{0mm}{0pt}
   $U(1)_Y$ \ & 1/6 & $-2/3$ & 1/3 & $-1/2$ & 1 & 0 & 1/2 & $- 1/2$ \\
    \hline
 \end{tabular}
 }
  \caption{The matter content of our D-brane model.  $i$ counts the 3  generations}
 \end{center}
\end{table}
\noindent
shown in Table 1, except for an extended Higgs sector (with 2 pairs per generation): 

\vspace{-4mm}

\noindent
More details of this proposal are given in section 4.
In section 5, we present a concrete geometric 
recipe for obtaining a compact CY manifold with all the required 
properties.

\newcommand{\ccc}{{\mbox{\large $c$}}}

\newcommand{\qq}{{\mathsf q}}
\newcommand{\pp}{{\mathsf p}}
\newcommand{\rr}{{\bf  r}}

\bigskip

\bigskip
\medskip

\noindent
\newsection{ General Strategy}

\medskip

We begin with a summary our general approach to string phenomenology. 
In subsection 2.1, we give
a quick recap of some relevant properties of D-branes at singularities. The reader
familiar with this technology may wish to skip to subsection 2.2. 
\medskip

\noindent
\newsubsection{D-branes at a CY singularity}

D-branes near Calabi--Yau singularities typically
split up into so-called fractional branes. Fractional branes can be thought of as particular bound state
combinations of D-branes, that wrap cycles of the local geometry.
In terms of the world-sheet CFT, 
they are in one-to-one correspondence with allowed conformally
invariant open string boundary conditions. Alternatively, by extrapolating to a large volume
perspective, fractional branes may be represented geometrically as particular well-chosen
collections of sheaves,  supported on corresponding
submanifolds within the local Calabi--Yau singularity.
For most of our discussion, however,
we will not need this abstract mathematical description;
the basic properties that we will use
have relatively simple topological specifications.

\newcommand{\is}{\! &\! = \! &\!}

There are many types of CY-singularities, and some are, in principle, good candidates for
finding realistic D-brane gauge theories. For concreteness, however, we
specialize to the subclass of singularities which are asymptotic to
a complex cone over
a del Pezzo surface $X$.  D-brane theories on del Pezzo singularities
have been studied in \cite{MH,martijn,chris}.

A del Pezzo surface is a manifold of complex
dimension 2, with a positive first Chern class.
Each
del Pezzo surface other than $\PP^1 \times \PP^1$
can be  represented as $\PP^2$ blown up at $n \leq 8$ generic points;
such a surface is denoted by ${dP_n}$ and sometimes called ``the $n$-th
del Pezzo surface''.\footnote{This terminology is unfortunately at odds
with the fact that, for $n\ge 5$, ${dP_n}$ is not unique but actually
has $2n-8$ complex moduli represented by the location of the points.}
By placing an appropriate complex line bundle (the 
``anti-canonical bundle'') over $X={dP_n}$,
one obtains a smooth non-compact Calabi--Yau threefold.
If we then shrink the zero section of the line bundle to a point,
we get a cone over $X$, which we will call the conical del
Pezzo $n$ singularity and denote by $Y_{0}$.  (More general del Pezzo
singularities are asymptotic to $Y_0$ near the singular point.)
To specify the geometry of $Y_{0}$,
let $ds_X^2 = h_{a\bar{b}}dz^a dz^{\bar{b}}$ be a K\"ahler-Einstein metric over
the base $X$ with $R_{a\bar{b}} = 6 h_{a\bar{b}}$ and first Chern class
$\omega_{a\bar{b}} = 6 i R_{a\bar{b}}$.
Introduce the one-form $\eta = {1\over 3} d\psi + \sigma$ where $\sigma$ is defined
by $d\sigma = 2\omega$ and $0<\psi<2\pi$ is the angular coordinate for a circle bundle
over the del Pezzo surface. The Calabi--Yau metric can then be written as follows
\be
\label{one}
ds^2_{Y} = dr^2 + r^2\eta^2 + r^2 ds^2_X 
\ee
For the non-compact cone, the $r$-coordinate has infinite range.
Alternatively, we can think of the del Pezzo singularity as a localized region
within a compact CY manifold, with $r$ being the local radial coordinate distance from the singularity.
We will consider both cases.
\\[3mm]
$
\begin{array}{cc}
 \noindent
  \hspace{-2mm}
\parbox{9.3cm}{
\addtolength{\baselineskip}{1mm}
${}$ \ \
The del Pezzo surface $X$ forms a four-cycle within the
full three-manifold $Y$, and itself supports several non-trivial
two-cycles.
Now, if we consider IIB string theory on a del Pezzo
singularity, we should expect to find a basis
of fractional branes that spans the complete
homology of $X$: the del Pezzo 4-cycle itself may be
wrapped by any number of D7-branes, any 2-cycle within $X$
may be wrapped by one or more D5 branes, and the point-like
D3-branes occupy the 0-cycle within $X$.
The allowed fractional
 branes, however,  typically do not correspond to single
branes wrapped on some given cycle, but rather }
\parbox{7cm}{\ \ \ \ \ \ \ \
\leavevmode\hbox{\epsfxsize=6cm \epsffile{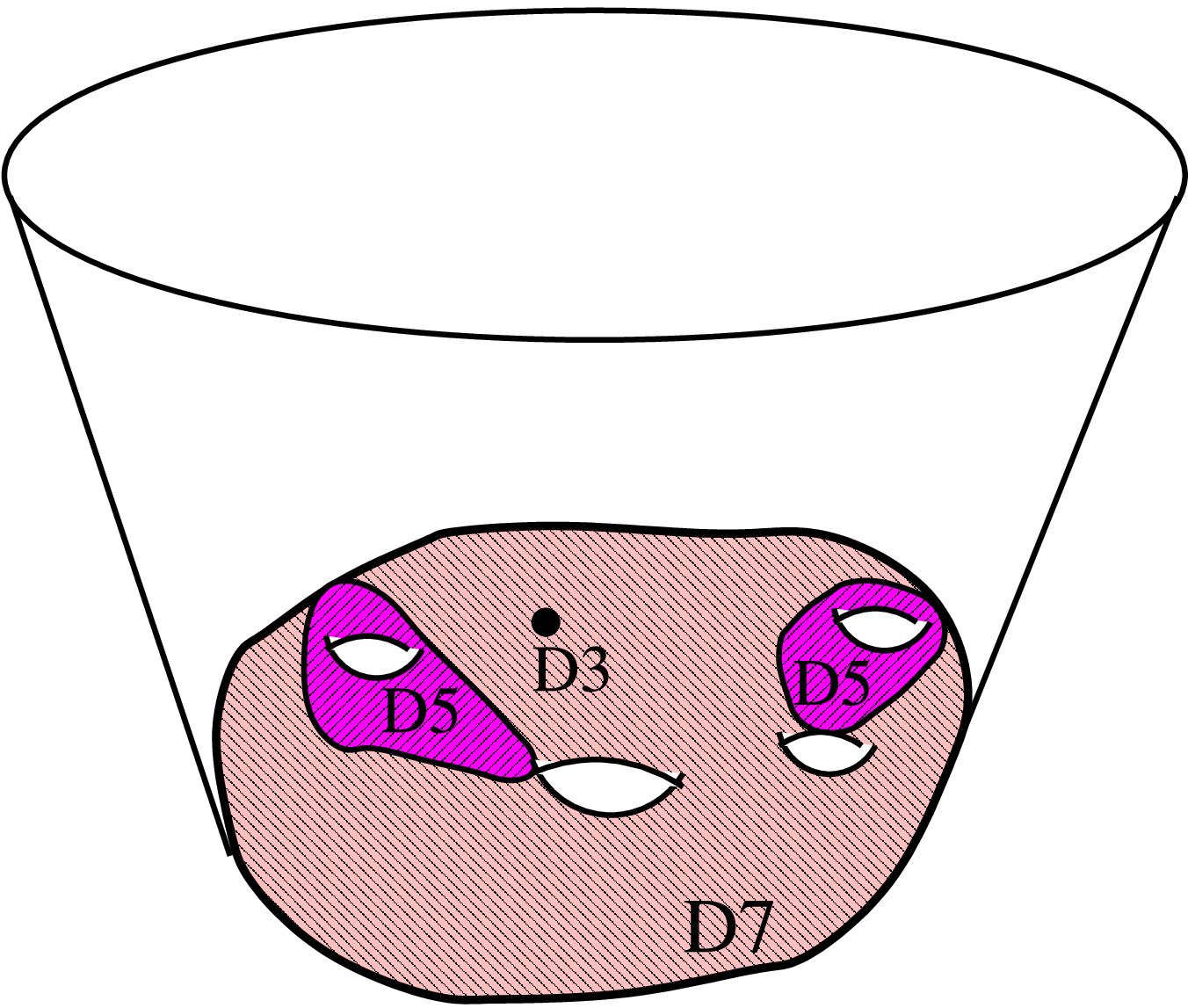}}
}
\end{array}
$\\[2mm]
to particular bound states $\FF_\sss$, each characterized by a charge vector of the form
\be
{\rm ch}(\FF_\sss)= (\, r_\sss, \, p_{s}^{\,\aaa}, \, q_s \, )
\ee
Here $r_\sss\! = {\rm rk}(\FF_\sss)$ is the rank of the fractional brane $\FF_s$,
and is equal to the D7-brane wrapping number around $X$.
The number $q_s\! =\! {\rm ch}_2(\FF_\sss)$ is the 2nd Chern character
of $\FF_\sss$ and counts the D3-brane charge. Finally, the integers  $p_s^{\, \aaa}$
are extracted from the first  Chern class of $\FF_\sss$ via
\be
p_s^\aaa = \int_{\alpha_\aaa}\! c_1(\FF_\sss) 
\ee
where $\alpha_\aaa$ denotes an integral basis of $H_2(X)$.
Geometrically, $p_s^{\, \aaa}$ counts the number of times the D5-brane component of
$\FF_\sss$ wraps the 2-cycle $\alpha_\aaa$.

\medskip

For a given geometric singularity, it is a non-trivial problem to find consistent
bases of fractional branes that satisfy all geometric stability conditions.
For del Pezzo singularities, a special class of consistent bases 
are known, in the form of so-called
exceptional collections \cite{karpov, martijn, chris}.
These satisfy special properties, that in particular
ensures the absence of adjoint matter in the world-volume gauge theory, besides
the gauge multiplet. The formula for the intersection product between two
fractional branes $\FF_i$ and $\FF_j$ of an exceptional collection reads
\be
\label{pair}
\#(\FF_i,\FF_j) = {\rm rk}(\FF_i) \, {\rm deg}(\FF_j) - {\rm rk}(\FF_j)\, {\rm deg}( \FF_i) \, \equiv \, \chi_{ij}
\ee
Here the degree of $\FF_i$ is given by
${\rm deg}( \FF_i)\! =\! - c_1(\FF_i)\! \cdot\!  K$ with $K$ the canonical class on~$X$. It 
equals the intersection number between
the D5 component  of $\FF_i$ and the del Pezzo surface. The intersection number $\chi_{ij}$ 
governs the
number of massless states of open strings that stretch between the two fractional branes $\FF_i$ and $\FF_j$.

The world-volume theory on D-branes near a CY singularity takes
the form of a quiver gauge theory. For exceptional collections, the rules 
for drawing the quiver diagram are: \footnote{These rules can be generalized by including orientifold planes
that intersect the CY singularity. We will elaborate on this possibility in the concluding section.}
 \ 
 (i)\, draw a single node for every basis element $\FF_i$ of the collection,
(ii) connect every pair of nodes with $\chi_{ij}
>0$ by an oriented line with multiplicity $\chi_{ij}$. Upon assigning
a multiplicity $n_i$ to each fractional brane $\FF_i$, one associates to the 
quiver diagram a quiver gauge theory.  
The gauge theory has a $U(|n_i|)$ gauge group factor for every node $\FF_i$, as well as
$\chi_{ij}$ chiral multiplets in the bi-fundamental representation $(n_i, \bar{n}_j)$.
The multiplicities $n_i$ can be freely adjusted, provided the resulting
world volume theory is a consistent ${\cal N} \! =\! 1$ gauge theory, free of any non-abelian
gauge anomalies.

Absence of non-abelian gauge anomalies is ensured if
at any given node, the total number of incoming and outgoing lines (each weighted by
the rank of  the gauge group at the other end of the line)  are equal:
\be
\sum_j \chi_{ij} \, n_j = 0\, .
\ee
This condition is automatically satisfied if the configuration of
fractional branes constitute a single D3-brane, in which case the multiplicities
$n_i$ are such that $\sum_i n_i \, {\rm ch}(\FF_i)\! = \! (\, 0, 0, 1\, )$.
In general, however, one could allow for more general configurations,
for which the charge vectors add up to some non-trivial fractional brane charge.

For a given type of singularity, the choice of exceptional collection is not 
unique.\footnote{Each collection corresponds to a particular set of
stability conditions on branes, and determines a region in K\"ahler 
moduli space where it is valid.} Different
choices are related via simple basis transformations, known as mutations \cite{karpov}. However,
only a subset of all exceptional collections, that can be reached via mutations, lead to
consistent world-volume gauge theories.
The special mutations that act within the subset of physically relevant collections
all take the form of Seiberg dualities \cite{martijn,chris}.
Which of the Seiberg dual descriptions is appropriate is determined by the value of
the geometric moduli that determine the gauge theory couplings.

\bigskip

\noindent
\newsubsection{Symmetry breaking towards the SSM}

To find string realizations of SM-like theories we now proceed in two steps. First we look 
for CY singularities and brane 
configurations, such that the quiver gauge theory is just rich enough to contain the SM gauge
group and matter content. Then we look for a well-chosen symmetry breaking process that 
reduces the gauge group and matter content to that of the Standard Model, or at least
realistically close to it. 
When the CY singularity is not isolated, the moduli space of vacua for 
the D-brane theory has several components \cite{MoPl},
and the symmetry breaking we need is found on a component in which some of
the fractional branes move off of the primary singular point along
a curve of singularities (and other branes are replaced by appropriate
bound states).  This geometric insight into the symmetry
breaking allows us to identify an appropriate CY singularity, such that
the corresponding D-brane theory 
looks like the SSM.

The above procedure was used in \cite{MH} to construct a semi-realistic theory
from a single D3-brane on a partially resolved del Pezzo 8 singularity (see also section 4). 
The final model of \cite{MH},
however, still has several extra $U(1)$ factors besides the hypercharge symmetry.
Such extra $U(1)$'s are characteristic of D-brane constructions: typically, one
obtains one such factor for every fractional brane. As will be explained in what follows, 
whether or not these extra $U(1)$'s actually survive as massless gauge symmetries 
depends on the topology of how the singularity is embedded inside of a 
compact CY geometry.  

In a
string compactification, $U(1)$ gauge bosons may acquire a non-zero
mass via coupling to closed string RR-form fields. We will describe this mechanism
in some detail in the next section, where we will show that  the $U(1)$ bosons that remain
massless are in one-to-one correspondence with 2-cycles, that are {\it non-trivial}
within the local CY singularity but are {\it trivial} within the full CY threefold.
This insight in principle makes it possible to ensure -- via the topology of the CY compactification -- 
that, among all $U(1)$ factors of the D-brane gauge theory, only the hypercharge survives
as a massless gauge symmetry.

The interrelation between the 2-cohomology of the del Pezzo base of the singularity, and
the full CY threefold has other relevant consequences.
Locally, all gauge invariant couplings of the D-brane theory can be varied  via
corresponding deformations of the local geometry. This local tunability is one of the
central motivations for the bottom-up approach to string phenomenology. The embedding into a
full string compactification, however, typically introduces a topological obstruction against
varying all local couplings: only those couplings that descend from moduli of the full CY survive.
Their value will need to be fixed via a dynamical moduli stabilisation mechanism.

 \bigskip

 \noindent
 \newsubsection{Summary}

Let us summarize our general strategy in terms of a systematized set of steps:\\[2mm]
\noindent
(i) Choose a non-compact CY singularity, $Y_0$,
and find a suitable basis of fractional branes $\FF_i$ on it.
Assign multiplicities $n_i$ to each  $\FF_i$ and enumerate the resulting quiver gauge theories.\\[1.5mm]
(ii) Look for quiver theories that, after symmetry breaking, produce an SM-like theory.
Use the geometric dictionary to identify the corresponding (non-isolated) CY singularity.\\[1.5mm]
(iii) Identify the topological condition that isolates hypercharge as the only massless $U(1)$.
Look for a compact CY threefold, with the right topological properties, that contains ${Y}_0$.
\medskip

\noindent
In principle, it should be possible to automatize all three of these steps and thus set up
a computer-aided search of SM-like gauge theories based on D-branes at CY singularities.

\bigskip

\bigskip

\bigskip

\noindent
\newsection{{$U(1)$ Masses via RR-couplings}}

The quiver theory of a D-brane near a CY singularity typically contains
several $U(1)$-factors, one for each fractional brane. Some of these
$U(1)$ vector bosons remain massless, all others either acquire a St\"uckelberg
mass via the coupling to the RR-form fields or get a mass through the Higgs mechanism \cite{ALE,mass,Antoniadis:2002,louisCY}.
We will now discuss the St\"uckelberg mechanism in some detail.

\noindent
\newsubsection{The $U(1)$ hypermultiplet}


To set notation, we first consider the $U(1)$ gauge sector on
a single fractional brane. 
Let us introduce the two complex variables
\be
\label{six}
\tau ={\theta \over 2\pi} + i {4\pi \over g^2}\, , \qquad \qquad
\quad S=\rho+i\zeta\, .
\ee
Here $\tau$ is the usual $SL(2, \Z)$ covariant complex coupling, that
governs the kinetic terms of the $U(1)$ gauge boson via (omitting fermionic terms)
\be
{\rm Im} \int \! d^2 \theta\,{\tau \over 8\pi }  \, W_{\alpha} W^\alpha = - {1\over 4g^2}\,
F\wedge * F + {\theta \over 32 \pi^2}\, F \wedge F
\ee
The field $S$ in (\ref{six})  combines a St\"uckelberg field $\rho$ and
a Faillet-Iliopoulos parameter $\zeta$. After promoting $S$
to a chiral superfield, we can write a
supersymmetric gauge invariant mass term for the gauge field via
\cite{ALE}
\be
\label{massterm}
\int \! d^4\te\, \frac{1}{4}({\rm Im}(S\! -\! \bar{S}\! - 2V))^2
=\frac{1}{2}(A - d \rho)\! \wedge  * (A -d\rho)\; -\, \zeta  D.
\ee
Here $D$ denotes the auxiliary field of the vector multiplet $V$.
Together with the mass term, we observe a Faillet-Iliopoulos term
proportional to $\zeta$.

The complex parametrization (\ref{six}) of the D-brane
couplings naturally follows from its embedding in type IIB string theory.
Without D-branes, IIB supergravity  on a Calabi--Yau threefold
preserves $\mathcal{N}\! =\! 2$ supersymmetry.
Closed string fields thus organize in  $\mathcal{N}\! =\! 2$ multiplets
\cite{Aspinwall, VafaWitten}.
The four real variables in (\ref{six}) all fit together as the scalar components
of a single hypermultiplet that appears after dualizing two components
of the so called double tensor multiplet \cite{Grimm}.
Since adding a D-brane breaks half the supersymmetry, the hypermultiplet
splits into two complex $\mathcal{N}\! =\! 1$ superfields with  scalar
components $\tau$ and $S$. 
The hypermultiplet of a single D3-brane derives directly from the
10-d fields, via 
\ba
\label{tau}
\tau \is C_{0} + i e^{-\phi}, \nonumber \\[3mm] 
d S\, \is * d (C_{\it 2} + \tau B_{\it 2})
\ea
A $dP_n$ singularity $Y_0$ supports a total of $n+3$ independent fractional branes, and
a typical D-brane theory on $Y_{0}$ thus contains
$n+3$ separate $U(1)$ gauge factors. In our geometric dictionary, we need to account for
a corresponding number of closed string hypermultiplets.

In spite their common descent from the hypermultiplet, from the
world volume perspective $\tau$ and $S$ appear to stand on somewhat different
footing: $\tau$ can be chosen as a non-dynamical coupling, whereas $S$ must
enter as a dynamical field.  In a decoupling limit, one would expect that all closed
string dynamics strictly separates from the open string dynamics on the brane,
and thus that all closed string fields freeze into fixed, non-dynamical couplings.
This decoupling can indeed be arranged,  provided the $U(1)$ symmetry is
non-anomalous and one starts from a D-brane on a {\it non-compact} CY singularity.
In this setting, $\tau$ becomes a fixed constant as expected, while $S$
completely decouples, simply because the $U(1)$ gauge boson stays massless.


\bigskip

\noindent
\newsubsection{Some notation}


As before, let $Y_0$ be a non-compact CY singularity given by a complex cone over 
a  base $X$. A complete basis of IIB fractional branes on $Y_0$ 
spans the space of compact, even-dimensional  homology cycles within $Y_{0}$, which
coincides with the even-dimensional homology of $X$. 
The 2-homology of the $n$-th del Pezzo surface $dP_n$ is
generated by the canonical class  $\alpha_0= k$, plus $n$ orthogonal
 2-cycles $\alpha_i$. 
Using the intersection pairing within
the threefold  $Y_{0}$,  we introduce the dual 4-cycles
$\beta^\bbb$ satisfying
\ba
\qquad
\alpha_\aaa \ccap \beta^\bbb = \delta_\aaa^\bbb
\qquad \quad \mbox{{\scriptsize A, B} {\footnotesize  $=0,\ldots, n$}}
\ea
The cycle $\beta^0$,
dual to the canonical class
$\alpha_0$, describes the class of the del Pezzo surface $X$ itself, and forms the
only compact 4-cycle within  $Y_{0}$.
The remaining $\beta$'s are all non-compact and extend in the radial direction
of the cone.
The degree zero two-cycles $\alpha_i$, that satsify $\alpha_0 \ccap \alpha_i = 0$, 
have the intersection form
\be
\alpha_i \ccap \alpha_j = - A_{ij} 
\ee
where $A_{ij}$ equals  minus the Cartan
matrix of $E_n$. The canonical class has self-intersection $9-n$. 
In the following we will use the intersection matrix
\ba
\eta_{\aaa\bbb} = \left( \begin{array}{cc} 9-n & 0 \\[2mm] 0 & -A_{ij}\end{array}\right)
\ea
and its inverse $\eta^{\aaa\bbb}$
to raise and lower {\scriptsize A}-indices. 

\vspace{9mm}

\noindent
\newsubsection{Brane action}

\newcommand{\sump}{\mbox{\large $\sum\limits_{\raisebox{.5mm}{\scriptsize $p$}}$}}


The 10-d IIB low energy field theory contains the  following bosonic
fields: 
the dilaton $\phi$, the NS 2-form $B$, the K\"ahler 2-form $J$ and the RR $p$-form potentials $C_p$,
with $p$ even from 0 to 8. (Note that the latter are an overcomplete set, since $dC_p = *_{{}_{10}}dC_{8-p}$.) From each of these fields, we can extract a 4-d scalar fields via integration over
a corresponding compact cycles within $Y_0$. 
These scalar fields parametrize the gauge invariant couplings of the D-brane theory. Near a CY singularity, however, $\alpha'$ corrections may be substantial, and this gauge theory/geometric 
dictionary is only partially under control. We will not attempt to solve this hard problem and will
instead adopt a large volume perspective, in which the local curvature is assumed to be small 
compared to the string scale. All expressions below are extracted from the leading order DBI action. Moreover, we drop all curvature contributions, as they do not affect the main conclusions.
To keep the formulas transparant, we omit factors of order 1 and work
in $\ell_s=1$ units. For a more precise treatment, we refer to \cite{louisCY}.

The D-brane world-volume theory lives on a collection of fractional
branes $\FF_s$, with properties as summarized in section 2.
Since the fractional branes all carry a non-zero D7 charge $r_\sss$, we
can think of them as D7-branes, wrapping the base $X$ of the
CY singularity $r_\sss$ times. We can thus identify the closed string couplings of
$\FF_\sss$ via its world volume action, given by the sum of a Born-Infeld
and Chern-Simons term via
\ba
\label{dbi}
\qquad \mathcal{S} =
\int\!
e^{-i^*_s \phi}\sqrt{{\rm det}(i_s^*\, (G +B) - F_s)} \,
+ \int\!
\sump\; i_s^* C_p \; e^{{F}_s - i_s^* B}\, .
\ea
Here $i^*_s$ denotes the pull-back of the various fields to the
world-volume of $\FF_s$;  it in particular encodes the information of the
D7 brane wrapping number $r_\sss$.
In case the D7 brane wrapping number $r_\sss$
is larger than one, we need to replace the abelian field 
strength $F_s$ to a non-abelian field-strength and take a trace where appropriate.
\footnote{
In general, there are curvature corrections to the DBI and Chern-Simons terms that would
need to be taken into account.
They have the effect of replacing the Chern character by \cite{Cheung:1997az}
${\rm Tr}(e^{F}) \to {\rm Tr}(e^{F})\sqrt{\hat{A}(T)\over \hat{A}(N)}$.
We will ignore these geometric contributions here, since they do not
affect the main line of argument.}

The D5 charges of $\FF_s$ are represented
by fluxes of the field strength $F_\sss$ through the various 2-cycles within $X$.
\ba
\label{pdef}
p_{s}^\aaa  = \int_{\alpha_\aaa} {\rm Tr} ( F_s)
\ea
Analogously, the D3-brane charge is identified with the instanton number charge.
\ba
\label{qdef}
q_s = \int_X {1\over 2} \, \! {\rm Tr}( F_s \wedge F_s)
\ea
The D5 charges $p_{s\aaa}$ are integers, whereas the D3 charge $q_s$ may take half-integer values.

The D-brane world volume action, since it depends on the field strength $F_s$ via the
combination $${\cal F}_s = F_s -i^*_s B\, ,$$ is invariant under gauge transformations
$B_{\it 2} \to  B_{\it 2} + d \Lambda$, $A_s \to  A_s + \Lambda$, with $\Lambda$
any one-form.
If $\Lambda$ is single valued, then the fluxes $p_s^\aaa$ of $F_s$ remain unchanged.
But the only restriction is that $d\Lambda$
belongs to an integral cohomology class on $Y$. The gauge transformations thus have
an integral version, that shifts the integral periods of $B$ into fluxes of $F_s$, and vice versa.
This integral gauge invariance naturally turns the
periods of $B_{\it 2}$ into angular variables.
The relevant $B$-periods for us are those along the 2-cycles of the del Pezzo surface $X$
\ba
b^\aaa \is \int_{\alpha_\aaa}\! B . 
\ea
The integral gauge transformations act on these periods and the D-brane charges via
\ba
\label{nshift}
\qquad b^\aaa & \to & b^\aaa + n^\aaa \nonumber \\[2mm]
\qquad p_s^\aaa & \to & p_{s}^\aaa + n^\aaa \, r_s\\[2mm]
\qquad q_s & \to & q_s - n_\aaa \, p_{s}^{\aaa}-{\textstyle{1\over 2}}r_sn_\aaa n^\aaa \nonumber
\ea
with $n^\aaa$ an a priori arbitrary set of integers. These transformations
can be used to restrict the $b^\aaa$ to the interval
between 0 and 1.

Physical observables should be invariant under (\ref{nshift}). This condition
provides a useful check on calculations, whenever done in a non-manifestly
invariant notation. A convenient way to preserve the invariance, is to introduce
a new type of charge vector for the fractional branes, obtained by
replacing in the definitions (\ref{pdef}) and (\ref{qdef}) the field strength $F_s$
by ${\cal F}_s$:
\be
{\mathsf {Q}}(\FF_s) = (\rr_s,\pp_{s\aaa}, \qq_s) \qquad
\ee
where $\rr_s = r_s$ and
\vspace{-8mm}
\begin{eqnarray}
\qquad \pp_{s\aaa}\! \! \is p_{s\aaa}- r_s b_\aaa 
\nonumber \\[4mm]
\qquad \qq_s\!  \is q_s +  p_{s}^{\, \aaa} \, b_\aaa- {\textstyle{1\over 2}}\, r_s \, b^\aaa b_\aaa
\end{eqnarray}
The new charges can take any real value, and are both invariant under (\ref{nshift}).

The charge vector is naturally combined into the central charge ${\mathsf Z}(\FF_s)$ of the fractional
brane $\FF_s$.  The central charge is an exact quantum property of the fractional brane, that can
be defined at the level of the worldsheet CFT as the complex number that  tells us which linear
combination of right- and left-moving supercharges the boundary state of the brane preserves.
It depends linearly on the charge vector:
\be
{\mathsf Z}(\FF_s) = {\mathsf \Pi} \cdot {\mathsf Q}(\FF_s),
\ee 
with ${\mathsf \Pi}$ some vector that depends on the geometry of the CY singularity.

In the large volume regime, one can show that
the central charge is given by the following expression:
\cite{minasian, Douglas:2000, kapustin}
\ba
 {\mathsf Z}(\FF_s) = \int_X \! e^{-i_s^*(B+iJ)}\, {\rm Tr}( e^{F_s}) 
\ea
where $J$ denotes the K\"ahler class on $Y$. Evaluating the integral gives
\ba
 \label{taus}
 {\mathsf Z}(\FF_s) \is \qq_s \, 
- \textstyle{1\over 2}\, \rr_s\, \zeta^\aaa\zeta_\aaa   - \, i\, \pp_{s \aaa} \, \zeta^\aaa
\, ,
\ea
with 
\vspace{-5mm}
\ba
\label{zeta}
\zeta^\aaa \is \int_{\alpha_\aaa}   J  \, .
\ea

 With this preparation, let us write the geometric expression for the couplings
 of the fractional brane $\FF_\sss$. From the central charge $Z(\FF_s)$, we can extract the effective
 gauge coupling via
 \ba
{4\pi \over g_s^2} = e^{-\phi} |{\mathsf Z}({\FF_s})|\, ,
\ea
which equals the brane tension of $\FF_s$. In the large volume limit, this relation directly
follows from the BI-form of the D7 world volume action. 
The phase of the central charge 
\ba
\zeta_s = {1\over \pi} {\rm Im} \log {\mathsf Z} (\FF_s)
\ea
gives rise to the
FI parameter of the 4-d gauge theory \cite{Douglas:2000}. 
Two fractional branes are mutually supersymmetric 
if the phases of their central charges are equal. Deviations of the relative phase generically
gives rise to D-term SUSY breaking, and such a deviation is therefore naturally interpreted as
an FI-term.

The couplings of the gauge fields to the RR-fields follows from expanding the CS-term of the action.
The $\theta$-angle reads
\ba
\theta_s \is \rr_s \theta_{{}_X} + \pp_{s\aaa} \theta^\aaa\, + \qq_s C_{\it 0}   , 
\ea
with
\ba
\theta^\aaa \is  \int_{\alpha_\aaa} \! C_{\it 2}\, , \qquad\qquad
\theta_{{}_X} = \int_X C_{\it 4}\, .
\ea
In addition, each fractional brane may support a St\"uckelberg field, which arises by dualizing  the RR  
2-form potential ${\mathcal C}_s$ that couples linearly to the gauge field strength via
\be
\label{linear}
{\mathcal C}_s \wedge F_s  
\ee
From the CS-term we read off that
\ba
\label{pot}
\quad {\mathcal C}_s \is \rr_\sss \, \ccc_{{}_X}
 \, + \, \pp_{s\aaa} \ccc^\aaa \,  +\,  \qq_s \, C_{\it 2}
\ea
\ba 
\label{ca}
\ccc^{\aaa}  \is \int_{\alpha_\aaa}\! C_{\it 4}\, .
\qquad \qquad \ccc_{{}_X} \, = \,  \int_X C_{\it 6}
\ea
Note that all above formulas for the closed string couplings all respect the integral
gauge symmetry (\ref{nshift}).

\bigskip

\bigskip

\noindent
\newsubsection{Some local and global considerations}

On $dP_n$ there are $n+3$ different fractional branes, with a priori as many independent gauge
couplings and FI parameters. However,
the  expressions (\ref{taus}) for the central charges ${\mathsf Z}(\FF_s)$  contain
only $2n+4$ independent continuous parameters: the dilaton, the (dualized) B-field, 
and a pair of periods ($b_\aaa$, $\zeta^\aaa$) for every of the $n+1$ 2-cycles in $dP_n$. 
We conclude that there must be two relations restricting the couplings. The gauge theory
interpretation of these relations is that
the $dP_n$ quiver gauge theory always contains two anomalous
$U(1)$ factors. As emphasized for instance in
\cite{Intriligator:2005aw}, the FI-parameters
associated with anomalous $U(1)$'s are not freely tunable, but
dynamically adjusted so that the associated D-term
equations are automatically satisfied. This adjustment relates
the anomalous FI variables and gauge couplings.

The non-compact cone $Y_0$ supports two compact
cycles for which the dual cycle is also compact, namely,  the canonical class and the
del Pezzo surface $X$. Correspondingly, we expect to find a normalizable
2-form and 4-form on $Y_0$.
\footnote{Using the form of the metric of the CY singularity as given in eqn (1), the normalizable 2-form can be found to be 
$\omega_X = {1\over r^4} \left[\omega - 2 \, { dr\over r}\! \wedge \eta \right]$. The normalizable 4-form is its Hodge dual. }
Their presence 
implies that two closed string modes survive as dynamical 4-d fields with normalizable
kinetic terms; these are the two axions $\theta^{\, 0}$ and $\theta_{{}_X}$ associated
with the two anomalous $U(1)$ factors. The two $U(1)$'s are dual to each other:
a $U(1)$ gauge rotation of one generates an additive shift in the $\theta$-angle of the other.
This naturally identifies the respective $\theta$-angles and St\"uckelberg fields via
\ba
\label{equal}
\theta^{\, 0} =\rho^{{}_X} \, , \qquad \qquad
\rho_{\, 0} = \theta_{{}_X} \, .
\ea
The geometric origin of these identifications is that the corresponding
branes wrap dual intersecting cycles\footnote{\small All other D5-brane components,
that wrap the degree zero cycles $\alpha_i$,
do not intersect any other branes within $Y$ (see formula \ref{pair}).
This correlates with the absence
of any other mixed $U(1)$ gauge anomalies.}. 

\newcommand{\eol}{\nonumber\\[2mm]}

We obtain non-normalizable harmonic forms on the non-compact cone $Y_{0}$
by extending the other harmonic 2-forms  $\omega_i$ on $X$ to $r$-independent
forms over  $Y_{0}$. The corresponding 4-d RR-modes are non-dynamical fields:
any space-time variation of $\ccc^\aaa$ with {\footnotesize A} $\neq 0$ would carry infinite kinetic
energy. This obstructs the introduction of the dual scalar field, the would-be
St\"uckelberg variabel $\rho_\aaa$, which would have a vanishing kinetic term.
We thus conclude that for the non-compact cone $Y_{0}$, all non-anomalous $U(1)$
factors remain massless. This is in accord with the expectation that in the non-compact
limit, all closed string dynamics decouples.\footnote{\small There is a slight subtlety, however.
Whereas the non-abelian gauge dynamics of a D-brane on a del Pezzo singularity
flows to a conformal fixed point in the IR, the $U(1)$ factors become infrared
free, while towards the UV, their couplings develop a Landau pole.  Via the holographic
dictionary, this suggests that the D-brane theory with non-zero $U(1)$ couplings
needs to be defined on a finite cone $Y_{0}$, with $r$ cut-off at some finite
value $\Lambda$. This subtlety will not affect the discussion of the compactied
setting, provided the location of all Landau poles is sufficiently larger than
the compactification scale.}

As we will show in the remainder of this section, the story changes for the compactified
setting, for D-branes at a del Pezzo singularity inside of a compact CY threefold $Y$. In this
case, a subclass of all harmonic forms on the cone $Y_{0}$ may extend to normalizable
harmonic forms on $Y$, and all corresponding closed string modes are dynamical
4-d fields. 

\bigskip
\bigskip

\noindent
\newsubsection{Bulk action}

The most general class of string compactifications, that may include the type of D-branes at 
singularities discussed here, is F-theory. For concreteness, however, we will consider
the sub-class of F-theory compactification that can be described by 
IIB string theory compactified on an orientifold CY threefold $Y = \widehat{Y}/\mathcal{O}$.
The orientifold map
$\mathcal{O}$ acts via
$$ 
\mathcal{O}=(-1)^{F_L}\Omega_p\sigma  
$$
where $F_L$ is left fermion number, $\Omega_p$ is
world-sheet parity,  and $\sigma$ is the involution acting on $Y$. It acts via its
pullback $\sigma^*$ on the various forms present. 
The fixed loci of $\sigma$ are orientifold planes.
We will assume that the orientifold planes do not
intersect the base $X$ of the del Pezzo singularity. 

\newcommand{\summa}{ \raisebox{.4mm}{$\sum\limits_a$}\; }

The orientifold projection eliminates one
half of the fields that were initially present on the full Calabi--Yau
space. Which fields survive the projection is determined by 
the dimensions of the corresponding even and odd cohomology space
$H^{(i,j)}_{+}$ and $H^{(i,j)}_{-}$ on
Calabi--Yau manifold $\widehat{Y}$.
Note that the orientifold projection in particular
eliminates the constant zero-mode components of $C_{\it 2}$, $C_{\it 6}$ and $B_{\it 2}$,
since the operator $(-1)^{F_L}\Omega_p$ inverts the sign of all these fields.

The RR sector fields give rise to
4-d fields via their decomposition into harmonic forms on $Y$, which we may identify
as elements of the $\bar\partial$ cohomology spaces $H^{(p,q)}$.
On the orientifold,  we need to decompose this space as
$H^{(p,q)}_{+}\oplus
H^{(p,q)}_{-}$,
where $\pm$ denotes the eigenvalue under the action of $\sigma^*$
\ba
\omega_{\alpha}   \in H^{(1,1)}_{+}(\widehat{Y},\Z)\, , \quad & &  \quad
\tilde{\omega}_{a}\in H^{(1,1)}_{-}(\widehat{Y},\Z)\, , \nonumber \\[3mm]
\omega^{\alpha}   \in H^{(2,2)}_{+}(\widehat{Y},\Z) \, ,\quad & &  \quad
\tilde{\omega}^{a}\in H^{(2,2)}_{-}(\widehat{Y},\Z) \, .\nonumber
\ea
The relevant RR fields, invariant under 
$\mathcal{O}=(-1)^{F_L}\Omega_p\sigma$, decompose as:
\ba
\label{c6}
C_{\it 2}\is 
 \theta^{a}(x)\,\tilde\omega_a\,  \nonumber 
 \\[3mm] C_{\it 4} \is 
 \ccc^{\, \alpha}(x)\, \omega_{\alpha} +
\rho_{\alpha}(x)\, {\omega}^{\alpha} \, ,\\[3mm]
 \ C_{\it 6}\, \is \,  {\ccc}_{a}(x)\, \tilde{\omega}^a \, .
\ea
Here $\ccc^{\alpha}$ and $\ccc_a$ 
are two-form fields
and $\rho_\alpha$ and $\theta^a$ are scalar fields.
Similarly, we can expand the K\"ahler form $J$ and NS B-field as\
\ba \label{b2}
 J\is\zeta^{\alpha}(x) \, \omega_{\alpha} \, , 
 \nonumber \\[3mm]
B_{\it 2} \is 
b^{a}(x) \, \tilde{\omega}_a\, .
\ea
We can choose the cohomology bases such that
\ba\label{condforms}
\int_Y\omega_{\alpha}\wedge{\omega}^{\beta}\is\delta^{\beta}_{\alpha},
\qquad \quad
\int_Y\tilde\omega_{a}\wedge\tilde{\omega}^{b}\, = \, 
\delta^{b}_{a}\, .
\ea
In what follows, $\omega_{{}_X}$ and $\tilde{\omega}_{{}_X}$ will denote the Poincar\'e dual
2-forms to the symmetric and anti-symmetric lift of ${X}$, respectively.

The IIB supergravity action
in string frame contains the following kinetic terms for the RR
$p$-form fields
\ba
\label{bulkaction}
\mathcal{S} \is
\int [ \, G^{ab}\, d\ccc_a\! \wedge\! *
d\ccc_b \, + \, 
G_{\alpha\beta}\, d\ccc^{\alpha} \! \wedge\! * d\ccc^{\beta}\, ]
\ea
where $G_{\alpha\beta}$ and $G^{ab}$
denote the natural metrics on the space of harmonic 2-forms on $Y$ 
\ba\label{Galhabeta}
G_{\alpha\beta}\is 
\int_Y\omega_{\alpha}\wedge*
\omega_{\beta}\, ,
\qquad \quad 
G^{ab}\, = \, 
\int_Y\tilde\omega^a\wedge* \tilde\omega^b \, .
\ea
The scalar RR fields $\theta^b$ and $\rho_\alpha$ are related to the above 2-form
fields via the duality relations:
\ba\label{cctildedual}
*\, d\theta^{\, b} \! \is \! -G^{ab} d \ccc_a \,  ,
\qquad \quad 
*\, d\rho_{\alpha} \, = \, 
G_{\alpha\beta} d\ccc^{\beta}\, .
\ea

The 4-d fields in (\ref{c6}) and (\ref{b2}) are all period integals of 10-d fields expanded in
harmonic forms. Each of the 10-d fields may also support a non-zero field
strength with some quantized flux. These fluxes play an important role in stabilizing
the various geometric moduli of the compactification. In the following we will assume that
a similar type of mechanism will generate a stabilizing potential for
all the above fields, that fixes their expectation values
and renders them massive at some high scale. The St\"uckelberg
and axion fields $\rho^\alpha$ and $\theta_0$ still play an important role in deriving the low energy
effective field theory, however.

\bigskip
\bigskip

\noindent
\newsubsection{Coupling brane and bulk}

Let us now discuss the coupling between the brane and bulk degrees of freedom.
A first observation, that will be important in what follows, is that
the harmonic forms on the compact CY manifold $Y$, when restricted to base $X$ of the
singularity, in general do not span the full cohomology of  $X$.  For instance, the
$2$-cohomology of $Y$ may have fewer generators than that
of $X$, in which case there must be one or more 2-cycles that are non-trivial within $X$
but trivial within $Y$. Conversely, $Y$ may have non-trivial cohomology elements
that restrict to trivial elements on $X$. The overlap matrices
\ba
\label{pi}
\Pi_{\alpha}^\aaa \is \int_{\alpha_\aaa} \omega_\alpha \, ,
\qquad \qquad 
\Pi^{\aaa}_a \, = \, \int_{\alpha_\aaa} \tilde\omega_a  \, ,
\ea
when viewed as linear maps between cohomology spaces $H^{(1,1)}(X,\Z)$ and
$H^{(1,1)}_{\pm}(Y,\Z)$, thus typically have both a non-zero kernel and cokernel.

As a geometric clarification, we note that the above linear map $\Pi$  between the 
2-cohomologies of $Y$ and $X$ naturally leads to an exact sequence
\begin{equation}\label{cohoexact}
...\to H^2(Y)\to H^2(X)\to H^3(Y/X)\to H^3(Y)\to H^3(X)\to ...
\end{equation}
where $X\subset Y$ is 
the 4-cycle wrapped by the del Pezzo in the CY 3-fold, $Y$.
The cohomology space $H^k(Y/X)$ is referred to as the
\lq relative' k-cohomology class. The map from $H^2(Y)$ to $H^2(X)$ in the exact
sequence is given by our projection matrix $\Pi$.  Since in our case 
$H^1(Y)\cong 0$ and $H^1(X) \cong 0$, we have from (\ref{cohoexact}):
\begin{equation}
\text{ker}[\Pi]\cong H^2(Y/X)
\end{equation}
or, in words, the kernel of our projection matrix is just the
relative 2-cohomology. Similarly, using the fact that $H^3(X)\cong 0$,
we deduce that
\begin{equation}
H^3(Y/X)\cong H^3(Y)\oplus\text{coker}[\Pi]
\end{equation}
In other words, the relative 3-cohomology $H^3(Y/X)$ is dual to the space of
all 3-cycles in $Y$ plus all 3-chains $\Gamma$ for which $\partial\Gamma\subset X$.

This incomplete overlap between the two cohomologies has immediate repercussions
for the D-brane gauge theory, since it implies that the compact embedding typically reduces
the space of gauge invariant couplings. The couplings are all period integrals of certain
harmonic forms, and any reduction of the associated cohomology spaces reduces the
number of allowed deformations of the gauge theory. This truncation is independent from
the issue of moduli stabilization, which is a {\it dynamical} mechanism for fixing
the couplings, whereas the mismatch of cohomologies amounts to a {\it topological}
obstruction.

By using the period matrices (\ref{pi}), we can expand the  topologically available
local couplings in terms of the global periods, defined in (\ref{c6}) and (\ref{b2}), as
\ba
b^{\, \aaa} = \Pi^\aaa_{\; a}\, b^a\, , \quad\ &  &\ \quad
\ccc^{\, \aaa} =  \Pi^{\aaa}_{\; \alpha}\,
\ccc^{\, \alpha}\nonumber \\[3mm]
 \theta^{\, \aaa} =  \Pi^{\aaa}_{\; a}\, \theta^{\, a} \, , \quad\ & &\ \quad
\zeta^{\, \aaa} =  \Pi^{\aaa}_{\; \alpha}\,
\zeta^{\alpha} \nonumber
\ea
By construction, the left hand-side are all elements of the subspace of $H^{(1,1)}$ that is
common to both $Y$ and $X$. The number of independent closed
string couplings of each type thus coincides with the rank of the corresponding overlap matrix.

\newcommand{\two}{{\mbox{\scriptsize 2}}}
\newcommand{\sums}{\mbox{\large $\sum\limits_{\raisebox{.5mm}{\scriptsize $s$}}$}}

As a special consequence, it may be possible to form
linear combinations of gauge fields $A_s$, for which the linear RR-coupling (\ref{linear})
identically vanishes. These correspond to linear combinations of $U(1)$ generators
$$
Q  =  \sums  k_s\, Q_s
$$
\vspace{-5mm}
such that
\ba
\sums \; k_s\, \rr_s\is 0\, ,    \qquad \qquad
\sums\; k_s\, 
\pp_{s \aaa}\, \Pi^{\aaa}_{\, \alpha}
\, = \, 
 0\, . 
\ea
The charge vector of the linear combination of fractional branes  $\sum_s k_s\, \FF_s$
adds up to that of a D5-brane wrapping a 2-cycle
within $X$ that is trivial within the total space $Y$. As a result, the corresponding
$U(1)$ vector  boson
$
A = \sum_s k_s \, A_s
$
decouples from the normalizable RR-modes,
and remains massless. 
This lesson will be applied in the next section.

Let us compute the non-zero masses.  Upon dualizing, or equivalently, integrating out the 2-form potentials, we obtain
the St\"uckelberg mass term for the vector bosons $A_s$
\ba
\label{finalmass} \quad
G_{{}_{\! XX} } \,\nabla \rho^{{}_X} \wedge * \nabla
\rho^{{}_X}
+ \, 
G^{\alpha\beta} \,
\nabla \rho_\alpha \wedge * \nabla \rho_\beta
\ea
\vspace{-4mm}
with
\ba
\nabla \rho^{{}_X} \is d \rho^{{}_X}
- \sums
\, \rr_{\sss} \, A_{s}\, , \nonumber \\[3mm]
\nabla \rho_\alpha \is d\rho_\alpha -
\sums \; \pp_{s\aaa} 
\, \Pi^\aaa_{\, \alpha}
\, A_s\, .
\ea
The vector boson mass matrix reads
\be
m^2_{s s'} =  G_{{}_{\! XX}}\, \rr_s \rr_{s'} +\, 
 G^{\alpha\beta}
\,\Pi^{\; \aaa}_{\alpha}
\Pi^{\; \bbb}_{\beta} \,
 \pp_{s \aaa} \pp_{s'\bbb}
\ee
and is of the order of the string scale (for string size compactifications).
It lifts all $U(1)$ vector bosons from the low energy spectrum, except for the ones
that correspond to fractional branes that wrap 2-cycles that are trivial within $Y$.
This is the central result of this section.

Besides via St\"uckelberg mass terms, vector bosons can also acquire a mass from
vacuum expectation values of charged scalar fields, triggered by turning on FI-parameters.
It is worth noting that for the same $U(1)$ factors for which the above
mass term (\ref{finalmass}) vanishes, the FI parameter cancels 
$$
\sums \, k_s\, \pp_{s \aaa}\Pi_\alpha^\aaa  \zeta^\alpha = 0
$$
 These $U(1)$ bosons thus remain massless, as long
as  supersymmetry remains unbroken.

\bigskip
\bigskip

\noindent
\newsection{SM-like Gauge Theory from a ${\rm dP_8}$ Singularity}

\medskip

We now apply the lessons of the previous section to the string construction of a Standard
Model-like theory of \cite{MH}, using the world volume theory of a D3-brane on a del Pezzo 8
singularity. Let us summarize the set up -- more details are found in \cite{MH}.

\bigskip
\bigskip

\newsubsection{A Standard Model D3-brane}

A del Pezzo 8 surface can be  represented as $\PP^2$ blown up at $8$ generic points.
It supports
nine independent
2-cycles: the hyperplane class $H$ in ${\PP}^2$ 
plus eight exceptional curves $E_i$  with intersection numbers $$
H\ccap H \, = \, 1, \ \ \  \quad E_i\ccap E_j =  - \delta_{ij} , \ \ \ \quad H \ccap E_i = 0\, .$$
The canonical class is identified as
$$K = -3 H +\mbox{$\sum_{i=1}^8$}\, E_i. $$
The degree zero sub-lattice of $H_2(X,\Z)$, the elements with zero intersection with
$c_1=-K$,  is  isomorphic to the root lattice of $E_8$.
The $8$ simple roots, all with self-intersection $-2$, can be chosen as 
\be
\label{ddd}
\alpha_i = E_i - E_{i+1},   \ \ \ \ \ \mbox{\footnotesize $i=1,\ldots, 7$}
\qquad \qquad \alpha_8 = h-E_1-E_2-E_3\, .
\ee
A del Pezzo 8 singularity thus accommodates 11 types of fractional branes $\FF_i$, which each are
characterized by charge vectors ${\rm ch}(\FF_i)$ 
that indicate their (D7, D5, D3) wrapping numbers.

Exceptional collections (=bases of fractional branes) on a del Pezzo 8 singularity have been constructed in \cite{karpov}.
 For  a given collection,
a D-brane configuration assigns multiplicity $n_i$ to each fractional brane $\FF_i$, consistent 
with local tadpole
conditions. The construction of \cite{MH} starts from a single
D3-brane; the multiplicities $n_i$ are such that the
charge vectors add up to $(0,0,1)$.
For the favorable basis of fractional branes described in \cite{MH}
(presumably corresponding to a specific stability region in K\"ahler
moduli space),
this leads to an ${\cal N}\!=\! 1$ quiver gauge theory with the gauge group
$
{\cal G}_0 = U(6) \times U(3) \times U(1)^9.
$
\footnote{This particular quiver theory is  related via a single Seiberg duality to the world volume theory of
a D3-brane near a  $\C^3/\Delta_{27}$ orbifold singularity -- the model considered earlier
in \cite{delta27} \cite{Aldazabal} as a possible starting point for a string realization of a SM-like
gauge theory.}
  
As  shown in \cite{MH}, this D3-brane quiver theory 
allows
a SUSY preserving symmetry breaking process to a semi-realistic
gauge theory with the gauge group
$$
{\cal G} =U(3) \times U(2) \times U(1)^7.
$$
The quiver diagram is drawn in fig 1. Each line represents three generations of
bi-fundamental fields.  The D-brane model thus has the same
non-abelian gauge symmetries, and the same quark and lepton content as the Standard Model.
It has an excess of Higgs fields -- two pairs per generation -- and several
extra $U(1)$-factors. We would like to apply the new insights obtained in the previous section
to move the model one step closer to reality, by eliminating all the extra $U(1)$ gauge
symmetries except hypercharge from the low energy theory.

To effectuate the symmetry breaking to ${\cal G}$, while preserving  ${\cal N}\! =\! 1$ supersymmetry, it
is necessary  turn on a suitable set of FI parameters and tune the superpotential $W$.\footnote{The superpotential $W$ 
contains Yukawa couplings for every closed oriented triangle in the  quiver diagram,
can be tuned via the complex structure moduli, in combination with suitable non-commutative 
deformations \cite{martijn2} of the del Pezzo surface.}
The D-term and F-term equations can then both be solved, while dictating expectation values
that result in the desired symmetry breaking pattern.
As first discussed in \cite{MoPl}\footnote{See also \cite{pol,dgm,ddg,lnv}.}
(in the context of $\mathbb{Z}_2\times\mathbb{Z}_2$ orbifolds),  when a
Calabi--Yau singularity is not isolated, the moduli space of D-branes
on that Calabi--Yau has more than one branch.  From a non-isolated
singularity, several curves $\Gamma_i$ of singularities will emanate,
each having a generic singularity type $R_i$, one of the ADE singularities.

In this non-isolated case, on one of the branches of the moduli space,
the branes move freely on the Calabi--Yau or its (partial) resolution,
and the FI parameters are identified with ``blowup modes'' which
specify how much blowing up is done.  But there are additional
branches of the moduli space associated with each $\Gamma_i$: on
such a branch, the FI
parameters which would normally be used to blow up the ADE singularity
$R_i$ are frozen to zero, and new parameters arise which correspond to
positions of $R_i$-fractional branes along $\Gamma_i$.  That is,
on this new branch, some of the $R_i$ fractional branes have moved
out along the curve $\Gamma_i$ and their positions give new parameters.

The strategy for producing the gauge theory of fig 1, essentially
following \cite{MH}, is this: by appropriately tuning the superpotential
(i.e., varying the complex structure) we can find a Calabi--Yau with
a non-isolated singularity---a curve $\Gamma$ of $A_2$ singular points---such 
that the classes $\alpha_1$ and
$\alpha_2$ have been blown down to an $A_2$ singularity on the (generalized)
del Pezzo surface where it meets the singular locus.\footnote{\small We 
will give an explicit description of a del Pezzo 8 surface with
the required $A_2$ singularity in the next section.}
Our symmetry-breaking
involves moving onto the $\Gamma$ branch in the moduli space, where
the $\alpha_1$ and $\alpha_2$ fractional brane classes are free to
move along the curve $\Gamma$ of $A_2$ singularities.  In particular,
these branes can be taken to be very far from the primary singular point
of interest, and become part of the bulk theory: any effect which they
have on the physics will occur at very high energy like the rest of
the bulk theory.

Making this choice removes the branes supported on $\alpha_1$ and
$\alpha_2$ from the original brane spectrum, and replaces other
branes in the spectrum by bound states which are independent of $\alpha_1$ and
$\alpha_2$.
The remaining bound state basis of the fractional branes obtained in \cite{MH}
is specified by the following set of charge vectors
\ba
\label{coll2}
 \begin{array}{ccc}
{\rm ch}(\FF_{1})  = (3, -2K + \! \mbox{$\sum\limits_{i=5}^8 E_i$}\! - E_4,  {\textstyle{1\over 2}} )
\\[3.5mm]
\, {\rm ch}(\FF_{2}) =\, (\, 3,\, \mbox{$\sum\limits_{i=5}^8 E_i$},  \textstyle{-2})\qquad \qquad \\[3.5mm]
{\rm ch}(\FF_{3})  = \,  (3,3H-\! \mbox{$\sum\limits_{i=1}^4 E_i$}, -\textstyle{ 1\over 2} ) \qquad
\end{array} 
\raisebox{-2mm}{$\begin{array}{ccc} & & \qquad
{\rm ch}({\FF}_{4})\, = \, (1, H -E_4, 0) \ \  \; \qquad \qquad \\[6mm]
& & \qquad \ \ \ \, {\rm ch}(\FF_{i})  \, = \, (1,- K\! + E_i , \,1\, ) \qquad \mbox{\scriptsize ${i =5,\, .  \,   ,8}$}\\[3.5mm]
& & \qquad \; {\rm ch}({\FF}_{9}) \, =  \, (1, 2H - {\mbox{$\sum\limits_{i=1}^4$}} E_i, 0) \qquad \qquad
\end{array}$}
\nonumber \\[-.8cm]
\ea
Here the first and third entry indicate the D7 and D3 charge; the second
entry gives the 2-cycle around wrapped by the D5-brane component of $\FF_i$.
As shown in \cite{MH}, the above collection of fractional branes is rigid, in the sense that
the branes have the minimum number of self-intersections and the corresponding gauge
theory is free of adjoint matter besides the gauge multiplet.
{}From the collection of charge vectors, one easily obtains the matrix of intersection
products via the fomula (\ref{pair}). 
One finds
\ba \label{chiminus}
\#(\FF_i,\FF_j) =\mbox{\footnotesize
$\left(\! \begin{array}{ccccccccc}
\ 0 & -3 & \ 0 & \ 1 & \ 1 & 1 & \ 1 & \ 1 & 1 \\
\ 3 &\ 0 &\ 3 & \ 2 & \ 2 & 2 & \ 2 & \ 2 & \ 2 \\
\ 0 & -3 & \ 0 & \ 1 & \ 1 & 1 & \ 1 & \ 1 & 1 \\
-1 & -2 & -1 & \ 0 & \ 0 & 0 & \ 0 & \ 0 & 0 \\
-1 & -2 & - 1 & \ 0 & \ 0 & 0 & \ 0 & \ 0 & 0 \\
-1 & -2 & -1 & \ 0 & \ 0 & 0 & \ 0 & \ 0 & 0 \\
-1 & -2 & -1 & \ 0 & \ 0 & 0 & \ 0 & \ 0 & 0 \\
-1 & -2 & -1 & \ 0 & \ 0 & 0 & \ 0 & \ 0 & 0 \\
-1 & -2 & -1 & \ 0 & \ 0 & 0 & \ 0 & \ 0 & 0
\end{array}\!
\right) $}\ea
which gives 
the quiver diagram drawn fig 1.  The rank of each gauge group corresponds to the
(absolute value of the)  multiplicity of the corresponding fractional brane, and has been
chosen such that weighted sum of charge vectors adds up to the charge of a single D3-brane.
In other words, the gauge theory of fig 1 arises from a
single D3-brane placed at the del Pezzo 8 singularity.

\begin{figure}[t]
\begin{center}
\leavevmode\hbox{\epsfxsize=8cm \epsffile{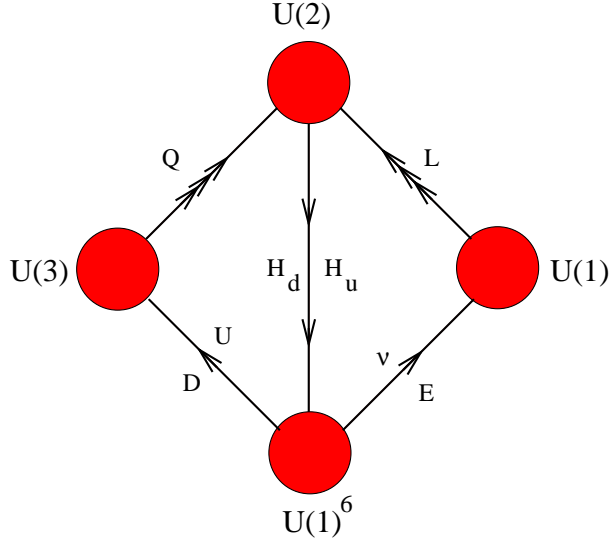}}
\caption{The MSSM-like quiver gauge theory obtained in \cite{MH}.  Each line represents three generations of bi-fundamentals. In the text below we will identify the geometric condition that isolates
the $U(1)_Y$ hypercharge as the only surviving massless $U(1)$ gauge symmetry.}
\end{center}
\end{figure}

Note that, as expected, all fractional branes in the basis (\ref{coll2}) have vanishing D5 wrapping
numbers around the two 2-cycles corresponding to the first two roots $\alpha_1$ and
$\alpha_2$ of $E_8$, since we have converted the FI parameters which
were blowup modes for those cycles into positions for $A_2$-fractional
branes.

After eliminating the two 2-cycles $\alpha_1$ and $\alpha_2$, the remaining
2-cohomology of the del Pezzo singularity is
spanned by the roots $\alpha_i$ with $i=3, .. , 8$ and the canoncial class $K$. 
Note that the total cohomology of the generalized del Pezzo surface with
an $A_2$ singularity is 9 dimensional,
and that the fractional branes (\ref{coll2}) thus form a complete basis.

\newcommand{\Ext}{{\rm Ext}}
\newcommand{\vQ}{\cal Q}
\newcommand{\vL}{\cal L}
\newcommand{\vbQ}{\overline{\vQ}}
\newcommand{\bu}{\overline u}
\newcommand{\bd}{\raisebox{-3pt}{$\overline d$}}
\newcommand{\bbe}{\overline e}
\newcommand{\bq}{\overline q}
\newcommand{\bH}{\overline H}
\newcommand{\bnu}{\overline \nu}



\bigskip
\bigskip

\newsubsection{Identification of hypercharge}

Let us turn to discuss the $U(1)$ factors in the quiver of fig 1, and identify the linear combination that defines hypercharge.  
We denote the node on the right by $U(1)_1$, and the overall $U(1)$-factors of the $U(2)$ and 
$U(3)$ nodes by $U(1)_2$ and $U(1)_3$, resp.  The $U(1)^6$ node at the bottom divides into
two nodes $U(1)^3_u$ and $U(1)^3_d$, where each $U(1)_u$ and $U(1)_d$ 
acts on the matter fields of
the corresponding generation only.  We denote the nine  $U(1)$ generators
by $\{Q_1, Q_2, Q_3, Q^i_u,Q^i_d, \}$. The total charge
$$
Q_{tot} = \sum_s Q_s
$$
decouples: none of the bi-fundamental fields is charged under $Q_{tot}$.
Of the remaining eight generators, two have mixed $U(1)$ anomalies. As discussed in
section 3, these are associated to fractional branes that intersect compact cycles
within the del Pezzo singularity. {In other words, any linear combination of charges
such that the corresponding fractional brane has zero rank and zero degree is free
of anomalies.}

Hypercharge is identified with the non-anomalous combination
\be
\label{hyper}
Q_Y =  {1\over 2}  Q_1 - {1\over 6} Q_3 - {1\over 2}
\Bigl( \,  \mbox{$\sum\limits_{{}_{i=1}}^{{}^{{}_3}}$}\,  Q_d^i -  \mbox{$\sum\limits_{{}_{i=1}}^{{}^{{}_3}}$} \, Q^i_u\Bigr)
 \ee
The other non-anomalous $U(1)$ charges are
\ba
{1\over 3} Q_3-{1\over 2} Q_1 \is B-L,
\ea
together with four independent abelian flavor symmetries of the form
\be
Q^{ij}_{u,d} = Q_{u}^i - Q_{u}^j,
\qquad \quad
Q^{ij}_b  = Q^i_b - Q^j_b .
 \ee
 We would like to ensure that, among all these charges, only the hypercharge survives as
 a low energy gauge symmetry.  From our study of the stringy St\"uckelberg mechanism, we
 now know that this can be achieved if we find a
CY embedding of the $dP_8$ geometry such that only the particular 2-cycle associated with $Q_Y$
represents a trivial homology class within the full CY threefold.
We will compute this 2-cycle momentarily.

Let us take a short look at the physical relevance of the extra $U(1)$ factors in the
quiver of fig 2. If unbroken, they forbid in particular all $\mu$-terms,  the
supersymmetric mass terms for the extra Higgs scalars.
In the concluding section 6, we return to discuss possible string mechanisms for breaking the 
extra $U(1)$'s. First we discuss how to make them all massive.

The linear sum (\ref{hyper}) of $U(1)$ charges that defines $Q_Y$,
selects a corresponding linear sum of fractional branes, which we may choose as follows\footnote{\small
With this equation we do not suggest any bound state formation of fractional
branes. Instead, we simply use it as an intermediate step in determining the cohomology
class of the linear combination of branes, whose $U(1)$ generators add up to $U(1)_Y$.}
\be
\label{sumf}
\FF_{\rm Y} = {1\over 2} \Bigl( \, \FF_{3} -  \FF_1 -\!\! \mbox{$\sum\limits_{{i=4,5,9}}$} \FF_i\, + \! \! \mbox{$\sum\limits_{{i=6,7,8}}$} \FF_i\, \Bigr)
\ee
A simple calculation gives that, at the level of the charge vectors
\be
{\rm ch}(\FF_{\rm Y} ) = (\,0\, , \, -\alpha_4,\; 
\mbox{\large ${1\over 2}$} \, ) \qquad \quad \alpha_4 = e_5-e_4
\ee
We read off that the 2-cycle associated with the hypercharge generator $Q_Y$ is the one represented by the simple root $\alpha_4$. 

\newcommand{\alphak}{{J}}

\begin{figure}[t]
\begin{center}
\leavevmode\hbox{\epsfxsize=9cm \epsffile{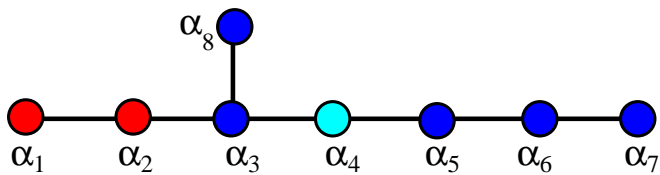}}
\caption{Our proposed D3-brane realization of the MSSM involves a $dP_8$ singularity
embedded inside a CY manifold, such that two of its 2-cycles, $\alpha_1$ and $\alpha_2$, develop an $A_2$ singularity which forms part of a curve of
$A_2$ singularities on the CY, 
and all remaining 2-cycles except $\alpha_4$ are non-trivial within the full CY.}
\end{center}
\end{figure}

We consider this an encouragingly simple result. Namely,
when added to the insights obtained in the previous section, we arrive at the following attractive
geometrical conclusion: we can ensure that all extra $U(1)$ factors except hypercharge
acquire a St\"uckelberg mass,  provided we can find  compact CY manifolds with a del
Pezzo 8 singularity, such that only $\alpha_4$ represents a trivial homology class.
Requiring non-triviality of all other 2-cycles except $\alpha_4$ not only helps with
eliminating the extra $U(1)$'s, but also keeps a maximal number of gauge invariant couplings
in play as dynamically tunable moduli of the compact geometry.
In particular, to accommodate the construction of the SM quiver theory of fig 1,
the complex structure moduli of the compact CY threefold must allow for the formation of
an $A_2$ singularity within the del Pezzo 8 geometry\footnote{
This can be done without any fine-tuning, as follows.
The complex structure of $Y$ is fixed 
via the GVW superpotential, which for given integer 3-form
fluxes 
takes the form $
W
 \! = \! (n_\alphak + \tau m_\alphak)\, \Omega^\alphak\,$ 
where $\Omega^\alphak$ denote the periods of the 3-form $\Omega$.
Now choose the integer fluxes to be invariant under the diffeomorphisms that
act like Weyl reflections in $\alpha_1$ and $\alpha_2$. $W$ then has
an extremum for $\Omega^\alphak$ invariant both Weyl reflections, which is the locus
where  $dP_8$ has the required $A_2$ singularity.}, with $\alpha_4$
representing a trivial cycle and all other cycles being nontrivial.
In the next section, we will present a general geometric  prescription for constructing 
a compact CY embedding of the $dP_8$ singularity  with all the desired topological 
properties.

\bigskip

\bigskip
\bigskip

\newsection{Constructing the Calabi--Yau threefold}

\medskip

\newcommand{\cO}{{\cal O}}

It is not difficult to find examples of compact CY threefolds that contain a $dP_8$ singularity.
Since a $dP_8$ surface can be constructed as a hypersurface of degree six in the weighted projective space $W\PP_{(1,1,2,3)}$, one natural route is to look among realizations of CY threefolds 
as hypersurfaces in weighted projective space, and identify coordinate regions where the
CY equation degenerates into that of a cone over $dP_8$. Examples of this type 
are the CY threefolds  obtained by resolving singularities of 
degree 18 hypersurfaces $W\PP_{(1,1,1,6,9)}$, considered in \cite{eighteen}.
This class of CY manifolds, however, has only two K\"ahler classes, and therefore can not 
satisfy our topological requirement that all 2-cycles of $dP_8$  
except $\alpha_4$ lift to non-trivial cycles within $Y$. On the other hand, this
example does illustrate the basic phenomenon of interest: since the 2-cohomology 
of $Y$ has only two generators, most 2-cycles within the $dP_8$
surface must in fact be trivial within $Y$.

A potentially more useful class of examples was recently considered in \cite{Diaconescu},
where it was shown how to 
construct a CY orientifold ${Y}$ as a  
$T^2$-fibration over any del Pezzo surface. The $T^2$ is represented in
hyperelliptic form, that is, as a two sheeted cover of a $\PP^1$. The 
$\PP^1$ fibration takes the form ${\PP}(\cO_X \oplus K_X)$ with 
$X$ the del Pezzo surface.
The covering space $\hat{Y}$ of $Y$ has a holomorphic involution $\sigma$,
which exchanges these two sheets, and the IIB orientifold on this CY surface is
obtained by implementing the projection $\mathcal{O}=(-1)^{F_L}\Omega_p\, \sigma.$
The ${\PP}^1$-fibration over the del Pezzo has two special
sections, $X_0$ and $X_\infty$, one of which can be contracted to a del Pezzo
singularity \cite{Diaconescu}. The total space of the fibration is the orientifold geometry $Y$. 
This set-up looks somewhat more promising for our purpose, since
all 2-cycles within $X$ are manifestly preserved as 2-cycles within the orientifold space $Y$. 
So a suitable modification the construction, so that only $\alpha_4$ is eliminated as a generator 
of $H^2(Y)$ while all other 2-cycles are kept,
would yield a concrete example of a CY orientifold with the desired global topology.
\footnote{A concrete proposal  is as follows. 
Tune the complex structure
so that the $dP_8$ has an automorphism which maps
$\alpha_4 \to - \alpha_4$, i.e. which acts as the Weyl reflection
$w(\alpha_4)$ on the homology lattice. One way 
to get such an automorphism is to let $X$ develop an $A_1$ singularity
with $\alpha_4$ as the $(-2)$ curve. The Weyl reflection then acts
trivially on the Calabi-Yau ${Y}$, but acts
non-trivially on the cohomology and the
string theory spectrum on $Y$.
We may then define a new holomorphic involution $\rho = w(\alpha_4) \circ
\sigma$ and consider the orientifold
$\mathcal{O}'=(-1)^{F_L}\Omega_p\,\rho.$
The $O7$-planes are at the same locus as
before, but the monodromy is slightly different. The
harmonic 2-form associated to $\alpha_4$ on $X$ still lifts to the cover space
$\widehat Y$, but as a generator of odd homology $H_-^{1,1}(\widehat{Y})$ instead of
$H_+^{1,1}(\widehat{Y})$. 
Therefore,  the FI-parameter and St\"uckelberg field associated to $\alpha_4$
are projected out, leading to a massless $U(1)_Y$.} 

Rather than following this route (of trying to find a specific 
compact CY manifold) we will instead give a general local prescription
for how to obtain a suitable compact embedding of the $dP_8$ singularity,
based only on the geometry of
the neighborhood of the singularity. This local perspective does not rely on 
detailed assumptions about the specific UV completion of the $dP_8$ model,
and thus combines well with  
our general bottom-up philosophy.

\bigskip
\medskip

\newsubsection{ Local Picard group of a CY singularity}

To begin, we discuss the local Picard group of a Calabi--Yau singularity,
and the effect it has on things such as deformations.

If $X$ is a (local or global) algebraic variety or complex analytic space
of complex dimension $d$, 
the {\em Weil divisors on $X$}, 
denoted $Z_{d-1}(X)$, are the $\mathbb{Z}$-linear combinations
of subvarieties of dimension $d-1$;
the {\em Cartier divisors on $X$}, denoted $\operatorname{Div}(X)$,
are divisors which are locally defined by a single equation $\{f=0\}$.
On a nonsingular variety, $Z_{d-1}(X)= \operatorname{Div}(X)$, so the
quotient group
\[Z_{d-1}(X)/\operatorname{Div}(X)\]
is one measurement of how singular the variety $X$ is.

The {\em principal Cartier divisors}\/ on $X$, denoted 
$\operatorname{Div}^0(X)$, are the divisors which can be written as the
difference of zeros and poles of a meromorphic function defined on all
of $X$, and the {\em Picard group of $X$}\/ is the quotient
\ba
\operatorname{Pic}(X)=\operatorname{Div}(X)/\operatorname{Div}^0(X). 
\nonumber 
\ea
If $X$ is sufficiently small, this is trivial, and one introduces a local
version of the group called the {\em local Picard group}: for a point $P\in X$,
\[ \operatorname{Pic}(X,P) = \operatornamewithlimits{lim}_\leftarrow Z_{d-1}(U)/\operatorname{Div}(U),\]
where the limit is taken over smaller and smaller open neighborhoods $U$
of $P$ in $X$.

Local Picard groups of Calabi--Yau singularities in complex dimesion $3$
were studied in detail
by Kawamata \cite{kawamata}, who showed that 
$\operatorname{Pic}(X,P)$ is finitely generated.  In our context,
we are mainly interested in the case where $X$ is a neighborhood
of a singular point $P\in X$ which is obtained by contracting a 
(generalized) del Pezzo surface $S$ in a Calabi--Yau space
$\widetilde{X}$ to a point via a map
$\pi:\widetilde{X}\to X$.\footnote{We allow $\widetilde{X}$ to have a curve of
 rational double point singularities, meeting $S$ in a rational double point,
which is why $S$ is called ``generalized'', following the terminology
of the mathematics literature.}  In this case, we can identify
$\operatorname{Pic}(X,P)$ with the image of the natural map
$\operatorname{Pic}(\widetilde{X}) \to \operatorname{Pic}(S)$.  
The rank of this image is always at least one: it follows from the
adjunction formula that there is always a divisor
$D_0$ on $\widetilde{X}$ such that $D_0+S$ is the divisor of a meromorphic
function on $X$, and the image of $D_0$ in $\operatorname{Pic}(S)$ is
the anticanonical divisor $-K_S$.

To take a simple, yet important example, suppose that $S=\mathbb{CP}^1\times
\mathbb{CP}^1\subset \widetilde{X}$ contracts to a Calabi--Yau singular point
$P\in X$.  There are two possibilities for 
$\operatorname{Pic}(X,P)$: it may happen that the two homology
classes $[\mathbb{CP}^1\times \{\text{point}\}]$ and 
$[\{\text{point}\}\times\mathbb{CP}^1]$ are the same in $H_2(\widetilde{X})$, 
in
which case $\operatorname{Pic}(X,P)\cong\mathbb{Z}$ (with the 
generator corresponding to $-K_S$), or it may happen that those two homology
classes are distinct, in which case 
$\operatorname{Pic}(X,P)\cong\mathbb{Z}^2$.  Note that if
$X$ is simply a cone over $S$, the classes will be distinct; on the 
other hand, the case $\operatorname{Pic}(X,P)\cong\mathbb{Z}$
is closely related to one of the key examples from Mori's
original pathbreaking paper \cite{mori} which started the modern classification
theory of algebraic threefolds.\footnote{In Mori's case, the normal
bundle of $S$ in $\widetilde{X}$ was $\mathcal{O}_S(-1,-1)$; on our
case, the normal bundle is $\mathcal{O}_S(-2,-2)$.}

The calculation of the local Picard group near a singular point
depends sensitively on the equation of the point.  Mori's example was
in fact a form of the familiar conifold singularity.  
It is common in the study of Calabi--Yau spaces to consider only the
``small'' blowups of such a singularity (which replace it by a 
$\mathbb{CP}^1$; however,
we could also choose to
simply blow up the singular point in the standard way, which would yield 
$\mathbb{CP}^1\times \mathbb{CP}^1$ with normal bundle 
$\mathcal{O}_{\mathbb{CP}^1\times \mathbb{CP}^1}(-1,-1)$.
The ``small'' blowups exist exactly when the two homology classes
$[\mathbb{CP}^1\times \{\text{point}\}]$ and
$[\{\text{point}\}\times\mathbb{CP}^1]$ are distinct; when they are the
same, we are in Mori's situation where small blowups do not exist.  How
do we determine this from the equation?
If we can write
the equation of the conifold singularity in the form
\begin{equation}\label{eq:conifold}
xy-zt=0
\end{equation}
then the two small blowups are obtained by blowing up the Weil divisors
$x=z=0$ or $x=t=0$, respectively.  However, if there are higher order
terms in the equation, the nicely factored form \eqref{eq:conifold}
may be destroyed:\footnote{The factored form can always be restored by
a local complex analytic change of coordinates, but that change of coordinates
may fail to extend over the entire Calabi--Yau.} this is Mori's case.
The case of a del Pezzo contraction, with 
normal bundle
$\mathcal{O}_{\mathbb{CP}^1\times \mathbb{CP}^1}(-2,-2)$,
is similar.\footnote{In that case, the small contractions would yield
a curve of $A_1$ singularities, as was crucial for the analysis of
\cite{morrison-seiberg}.}

If the neighborhood of the $\mathbb{CP}^1\times\mathbb{CP}^1$ is sufficiently 
large, the difference between the two cases can be detected by the topology
of the neighborhood.  When the two homology classes 
$[\mathbb{CP}^1\times \{\text{point}\}]$ and
$[\{\text{point}\}\times\mathbb{CP}^1]$ are the same, there is a $3$-chain
$\Gamma$ whose boundary is the difference between the two.  Such a $3$-chain
cannot exist if the two homology classes are distinct, so an analytic
change of coordinates which affects the factorizability of \eqref{eq:conifold}
will have the topological effect of creating or destroying such a 
$3$-chain $\Gamma$.\footnote{More details about the topology of this
situation can be found in \cite{greene-morrison-vafa}.}

\bigskip

\bigskip
\bigskip

\newsubsection{Construction of the CY threefold}

From this simple example, we can easily obtain more complicated ones,
including examples of the type we are interested in.  Let $S'$ be
a generalized del Pezzo surface obtained from $\mathbb{CP}^2$ by
(1) blowing up $5$ distinct points $P_4$, \dots, $P_8$ to curves 
$E_4$, \dots, $E_8$, (2) blowing up
a point $P_1$ and two points $P_2$ and $P_3$
infinitely near to $P_1$, (3) blowing down two out of the last
three exceptional
divisors to an $A_2$ singularity.
Note that the line $\ell_{45}$
through $P_4$ and $P_5$ lifts an an exceptional curve $E_{45}$, and that
the same del Pezzo surface $S'$ could be obtained starting from 
$\mathbb{CP}^1\times\mathbb{CP}^1$: in that case, one would blow up a point
$P_{45}$ to the curve $E_{45}$ observing that the two original
$\mathbb{CP}^1$'s which pass through $P_{45}$ lift to exceptional
curves $E_4$ and $E_5$, and then blowing up $P_6$, $P_7$, $P_8$, $P_1$,
$P_2$, $P_3$ as before.

\begin{figure}[t]
\begin{center}
\includegraphics[width=4.6in]{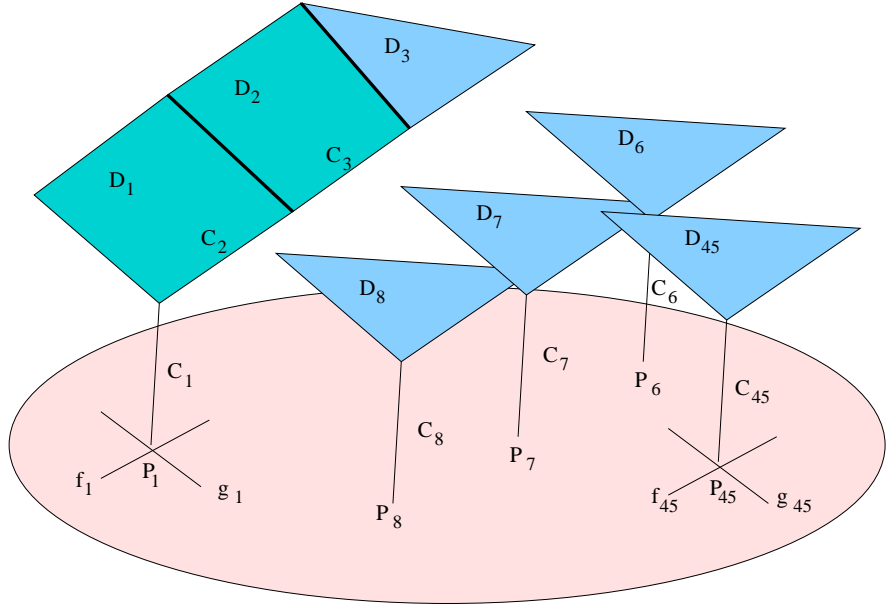}
\caption{Starting point of our construction of a CY threefold with the desired topology.  The curves $f_i$ and $g_i$ are fibers in the
two rulings on $S_1$.}
\label{figure3}
\end{center}
\end{figure}

We give an embedding into a Calabi--Yau in the following way.  Start with
$S_1:=\mathbb{CP}^1\times\mathbb{CP}^1$ embedded in a Calabi--Yau neighborhood
such that the two rulings are homologically equivalent in the Calabi--Yau.
We attach rational curves $C_{45}$, $C_6$, $C_7$, $C_8$ to the del Pezzo 
surface $S_1$,
meeting transversally at $P_{45}$, $P_6$, $P_7$ and $P_8$, and consider local
divisors $D_i$ meeting $C_i$ transversally at another point, for $i=45,6,7,8$.
We also attach a rational curve $C_1$ at $P_1$ which transversally meets the
first of a pair of ruled surfaces $D_1$ and $D_2$ which together can be
contracted to a curve of $A_2$ singularities.  We label the fiber of $D_1$'s
ruling which passes through $C_1\cap D_1$ by $C_2$, and we label
the fiber of $D_2$'s ruling which pass through $C_2\cap D_2$ by $C_3$.
We also consider a local divisor $D_3$ meeting $C_3$ transversally away
from its intersection with $C_2$.  This is all illustrated in 
figure~\ref{figure3}.

Each of the curves and surfaces we have used in this construction can
be embedded in a Calabi--Yau neighborhood, and those neighborhoods can
be glued together to form a Calabi--Yau neighborhood of the entire structure
illustrated in figure~\ref{figure3}.

\begin{figure}[hbtp]
\begin{center}
\includegraphics[width=4.8in]{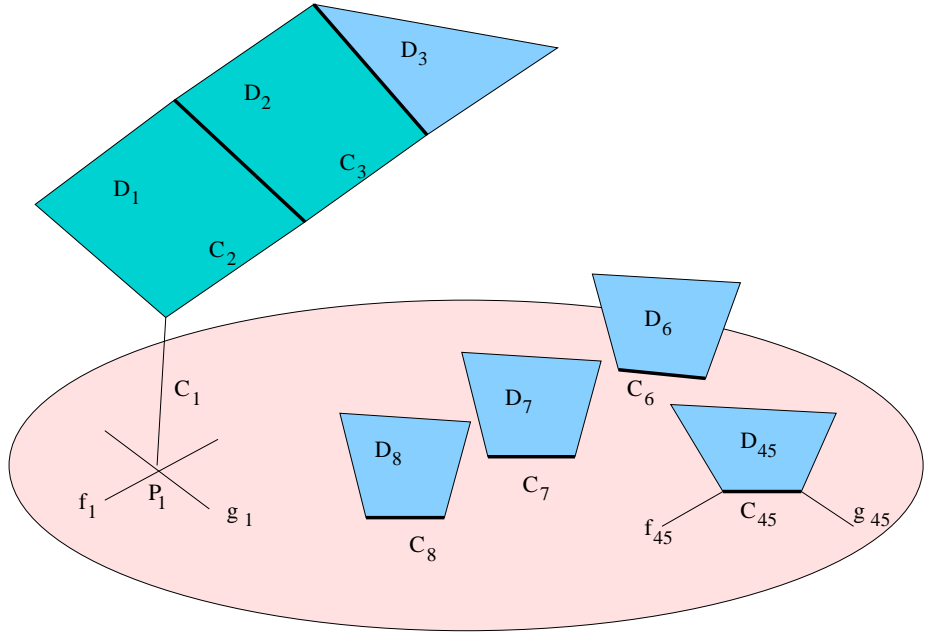}
\caption{The CY threefold after flopping the curves  $C_{45}, C_6, C_7$ and $C_8$.
The curves $f_{45}$, $g_{45}$, $C_{45}$, $C_6$, $C_7$, and $C_8$ are
all $(-1)$-curves on $S_5$.}  
\label{figure4}
\end{center}
\end{figure}

We now pass from this structure to the one we want by a sequence of flops.
First, we flop the curves $C_{45}$, $C_6$, $C_7$, and $C_8$, which has the 
effect
of blowing up $S_1$ at the four points $P_{45}$, $P_6$, $P_7$ and $P_8$
yielding a del Pezzo surface $S_5$.  The transformed surfaces
$D_{45}$, $D_6$, $D_7$, $D_8$ now meet $S_5$ in the flopped curves,
as indicated in figure~\ref{figure4}.

\begin{figure}[hbtp]
\begin{center}
\includegraphics[width=4.8in]{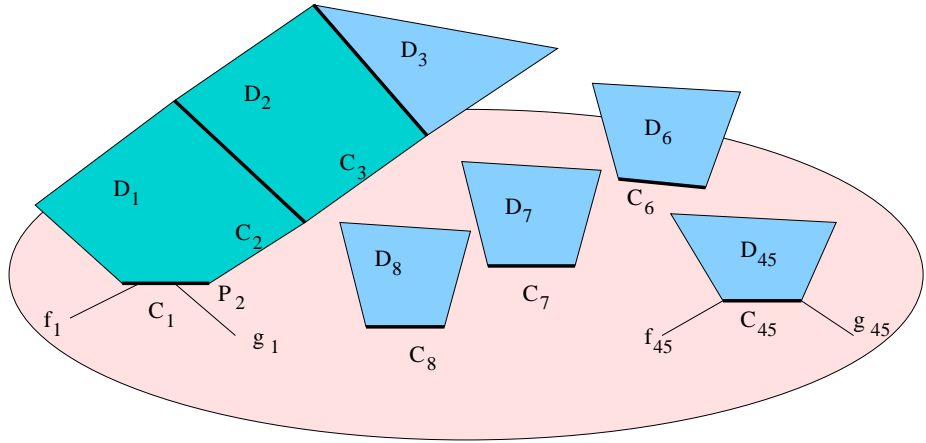}
\caption{The CY threefold after flopping $C_1$.
The curves $f_1$, $g_1$, and $C_1$ are additional $(-1)$-curves
on $S_6$.}
\label{figure5a}
\end{center}
\end{figure}

\begin{figure}[hbtp]
\begin{center}
\includegraphics[width=5in]{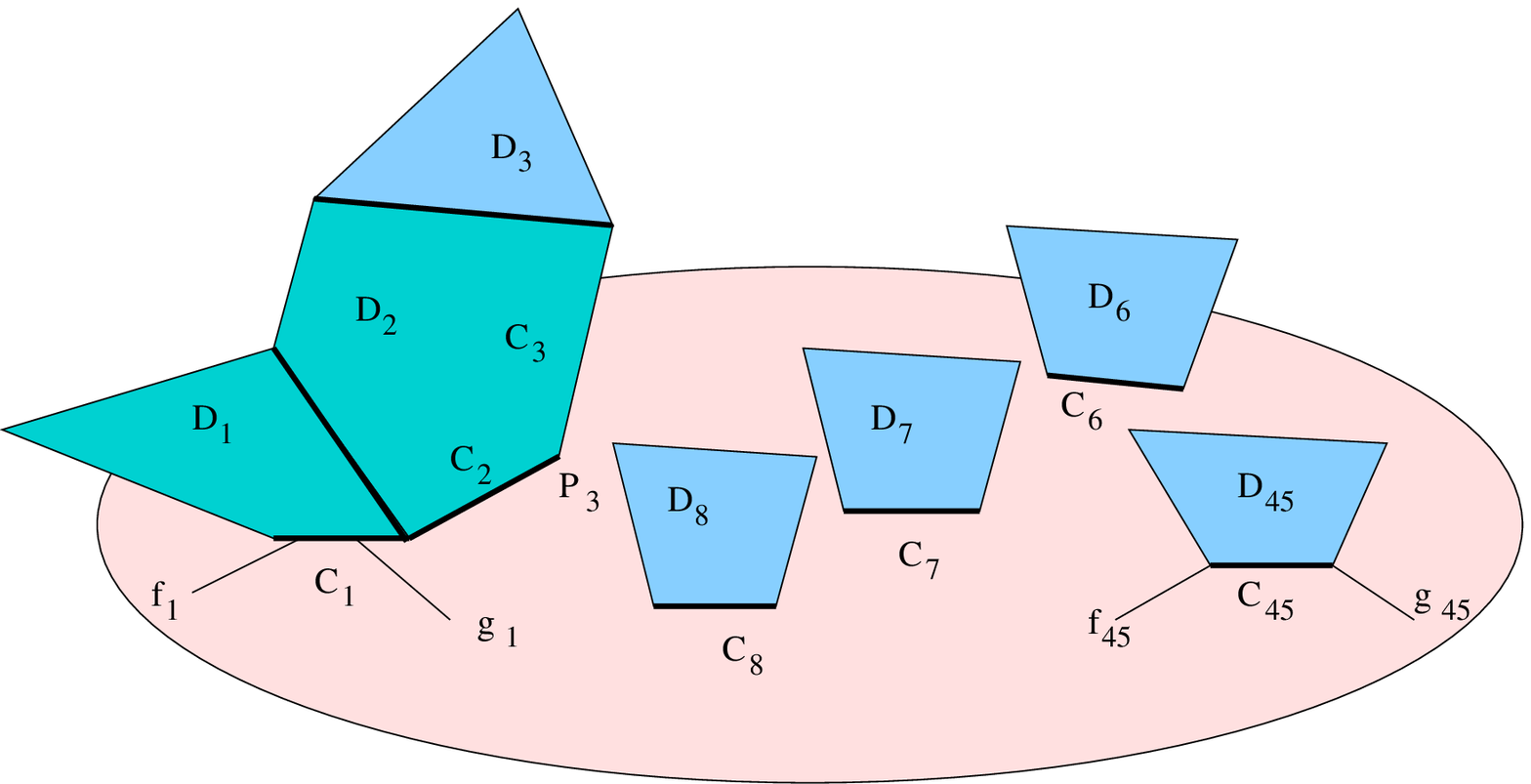}
\caption{The CY threefold after flopping $C_2$. 
The curve $C_1$ has become a $(-2)$-curve, and $C_2$ is a $(-1)$-curve
on $S_7$.}
\label{figure5b}
\end{center}
\end{figure}

Next, we flop the curve $C_1$, yielding a del Pezzo $S_6$ on which the point
$P_1$ has been blown up,
as indicated in figure~\ref{figure5a}.  The transformed surface $D_1$ meets $S_6$
in the flopped curve, and the transformed
curve $C_2$ meets $S_6$ in a point $P_2$ (``infinitely near'' to the first
point $P_1$). When $C_2$ is now
flopped, $S_6$ is blown up at $P_2$ to yield $S_7$, as indicated in figure~\ref{figure5b}.
The transformed surface $D_2$ meets $S_7$ in the most recently flopped curve.

\begin{figure}[hbtp]
\begin{center}
\includegraphics[width=5in]{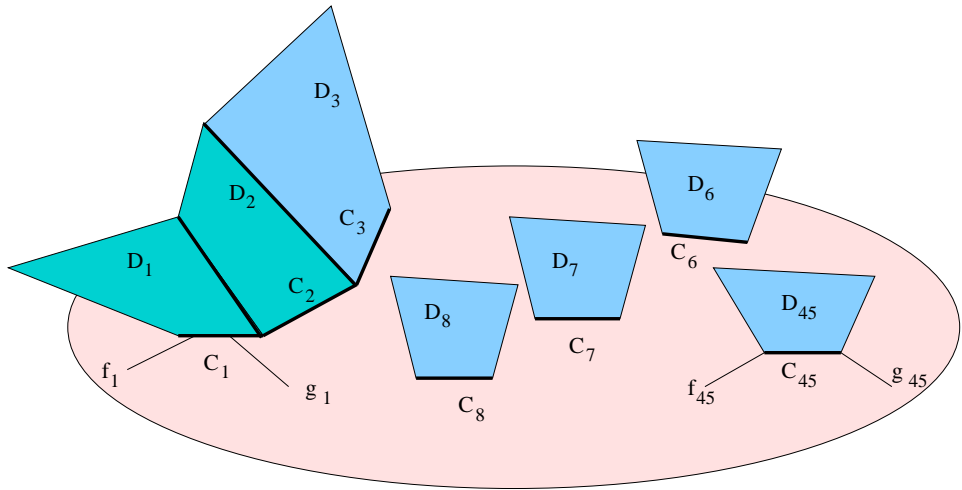}
\caption{The CY threefold after flopping $C_3$.
The curves $C_2$ has become a $(-2)$-curve, $C_1$ remains a 
$(-2)$-curve, and $C_3$ is
a $(-1)$-curve on $S_8$.}
\label{figure5c}
\end{center}
\end{figure}

\begin{figure}[hbtp]
\begin{center}
\includegraphics[width=5in]{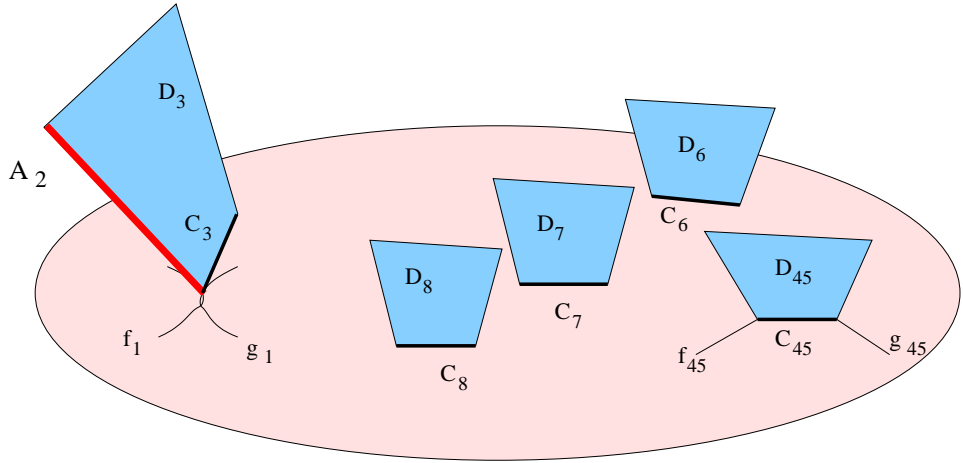}
\caption{The final configuration: a del Pezzo 8 surface with an $A_2$ singularity,
embedded into a Calabi--Yau such that two of its exceptional curves  are  
homologous.}
\label{figure6}
\end{center}
\end{figure}

The transformed curve $C_3$ meets $S_7$ in a point $P_3$  
(``infinitely near''
to $P_2$), and when $C_3$ is now flopped, $S_7$ is blown up at $P_3$
to yield $S_8$, as indicated in figure~\ref{figure5c}.  (The  
transformed surface
$D_3$
meets $S_8$ in the most recently flopped curve.)
To match the curves on $S_8$ to the standard basis for cohomology of a
del Pezzo, we set $e_1=C_1+C_2+C_3$, $e_2=C_2+C_3$, $e_3=C_3$, $e_4=f_ 
{45}$,
$e_5=g_{45}$, and $e_j=C_j$ for $j=6,7,8$ so that $\alpha_1=C_1$ and
$\alpha_2=C_2$.
We can now contract the
transforms of $D_1$ and $D_2$ to a curve of $A_2$ singularities,  
yielding
the configuration illustrated in figure~\ref{figure6}.

To achieve our final desired singular point, we contract the del Pezzo surface
$S'=S_8$ to a point.  Let us analyze the properties of this singular point.

First, it is not isolated: there is a curve of $A_2$ singularities which
eminates from our singular point.  This is one of the features we needed,
because it allows the fractional branes where were supported on $\alpha_1$
and $\alpha_2$ to move off of the singular point we are interested in,
into the bulk of the Calabi--Yau manifold.

Second, the local Picard group of this singular point has rank $6$: the
anticanonical divisor $D_0\equiv -K_S$ and the
transformed divisors $D_3$, $D_{45}$, $D_6$, $D_7$, $D_8$ generate
a subgroup of the local Picard group of rank six; 
if there were a seventh generator,
the map $\operatorname{Pic}(X)\to\operatorname{Pic}(S)$ would be surjective
and the original surface $S_1$ would have had the same property, so that
its local Picard group would have had rank $2$.  But
by construction, the surface $S_1$ had a local Picard group of rank $1$. 
We thus have demonstrated the 
presence of a 3-chain $\Gamma$, with boundary equal to the difference of two 
exceptional divisors. We can identify this difference with
$\alpha_4 = E_5-E_4$, which therefore does not exist as a homology class
in the full Calabi--Yau.

\medskip

Thus, via the above geometric procedure, we have succeeded in constructing
a compact CY threefold with the properties we need for our D-brane construction.
The outlined strategy furthermore preserves the main characteristics of our bottom-up 
perspective, since it only  refers to the local 
Calabi-Yau neighborhood of the singularity and does not  rely on unnecessary 
assumptions about the full string compactification.

An important physical assumption is that the compact embedding preserves the 
existence of all constituent fractional branes listed 
in eqn (\ref{coll2}). This is not entirely obvious, since, in particular, the D5 charge 
around the trivial cycle $\alpha_4$ is no longer a conserved quantum number:
one could imagine a tunneling process, in which the linear combination (\ref{sumf})
of fractional branes  combines into a single D5 wrapping $\alpha_4$,
which subsequently self-annihilates by unwrapping along the 3-chain $\Gamma$.
The tunneling process, however, is suppressed because it  is non-supersymmetric 
and the probability can be made exponentially small by ensuring 
that the 3-volume (measured in units of D-brane tension) of the 3-chain $\Gamma$ 
is large enough.

\bigskip

\bigskip

\bigskip

\noindent
\newsection{Conclusion and Outlook}

In this paper, we further developed the program advocated in \cite{MH}, aimed
at constructing realistic gauge theories  on the world-volume of D-branes at a
Calabi--Yau singularity. 
We have seen that several aspects of the world-volume gauge theory, such as the
spectrum of light $U(1)$ vector bosons and the number of freely tunable of couplings,
depend on the compact Calabi--Yau embedding of the
singularity. In section 3, we have worked out the stringy mechanism by which
$U(1)$ gauge symmetries get lifted. As a direct application of this result, we have shown
how to construct a supersymmetric Standard
Model, however with some extra Higgs fields, on a single D3-brane on a suitably
chosen Calabi--Yau threefold with a del Pezzo 8 singularity. 
The final result for the quiver gauge theory is given in fig 2, where in addition all extra 
$U(1)$ factors besides hypercharge are massive.

\bigskip

\bigskip

\newsubsection{$U(1)$ Breaking via D-instantons}

At low energies, the extra $U(1)$'s are  approximate global symmetries, which, if unbroken,
would in particular forbid $\mu$-terms. Fortunately, the geometry supports 
a plethora of D-instantons, that generically will break the $U(1)$ symmetries. Here we make some
basic comments on the generic form of the D-instanton contributions.

 The simplest type of D-instantons are the
euclidean D-branes that wrap compact cycles within the base $X$ of the CY singularity.
The `basic'  D-instantons of this type are in 1-1 correspondence with the space-time filling  fractional
branes: they are localized in ${\mathbb R}^4$, but otherwise
have the same Calabi-Yau boundary state  and preserve the same supersymmetries
as the fractional branes $\FF_s$. 
Apart from the exponential
factor 
$
  e^{-{8\pi^2/ g^2_{s}} \; +\,  i\, \theta_s},
$ 
their contribution is independent
of K\"ahler moduli and can thus be understood at large volume.
In this limit, the analysis has essentially already been done in \cite{Bershadsky:1996gx,Witten:1996bn,Ganor}.\footnote{For more recent discussions, see \cite{Blumenhagen:2006,Ibanez:2006,
FKMS}.}
The result agrees with the expected field theory answer \cite{Beasley:2004ys},
and sensitively depends on $N_c - N_f$, the number of colors minus  the number of flavors
 for the corresponding node.
In our gauge theory we have
$N_f > N_c$, in which case
the one-instanton contribution to the superpotential is of the schematic  form
%
\ba
 \delta W = {\Omega}(\Phi)
   \; e^{-8\pi^2 /g^2_{s} \; +\,  i\, \theta_s}
\ea
where $\Omega(\Phi)$ is a chiral multi-fermion operator \cite{Beasley:2004ys}.  The  theta angle $\theta_s$ in general contains an axion field, that is shifted by the 
anomalous $U(1)$ gauge rotations.
Instanton contributions to the effective action thus generally 
violate the anomalous $U(1)$ symmetries.

The story for the non-anomalous global $U(1)$ symmetries is analogous. 
The relevant
D-instanton contributions are generated by euclidean
D3-branes wrapping the dual 4-cycles $\Sigma_\alpha$ within $Y$.
The classical D-instanton action reads
$
S =  \mu_3  {\rm Vol}(\Sigma_\alpha)- i\!  \int_{\Sigma_\alpha} \! C_{\it 4}\
$
with $\mu_3$ the D3-brane tension.
Since $\int_{\Sigma_\alpha} \! C_{\it 4} = \rho_\alpha$ is the St\"uckelberg field,
we observe that the D3-instanton contribution to the superpotential takes the form
\ba
\label{kklt} \quad
\delta W 
= 
{\cal A}(\Phi)\; e^{-\mu_3 {\rm Vol}(\Sigma_\alpha) \; + \; i\rho_\alpha}\,
\ea
Here ${\cal A}(\Phi)$ denotes the perturbative pre-factor, the string analogue of
the fluctuation determinant, of the D-instanton.\footnote{
Note that, unlike all classical couplings, the D-instanton contributions (\ref{kklt})
are not governed by the local geometry of the singularity, but depend on the  
size of  dual cycles $\Sigma_\alpha$ that probe the full CY.  In fact,
eqn (\ref{kklt}) is a direct  generalization of the famous KKLT contribution to the 
superpotential, that
helps stabilize all geometric moduli of the compact Calabi--Yau manifold.}
Since the phase factor $e^{i\rho_\alpha}$
transforms non-trivially under the corresponding $U(1)$ rotation, 
the pre-factor 
must be oppositely charged. After gauge fixing,
the value of $\rho_\alpha$ will get fixed at by minimizing the potential, and $\rho_\alpha$ gets
lifted from the low energy spectrum. What remains is a superpotential term that, from the
low energy perspective, breaks the global $U(1)$ symmetry.

While we have not yet done the full analysis of these D-instanton effects,
 it  seems reasonable to assume that the desired $\mu$-terms can be generated 
via this mechanism.\footnote{In \cite{Berenstein:2006} it was argued
that $\mu$-terms can not arise in oriented quiver realizations of the SSM, like ours,
because they seem forbidden by chirality at the $SU(2)$ node, in 
case one would consider more than one single brane (so that $SU(2)$ 
 becomes $SU(2N)$).
It is important to note, however, that the form of the D-instanton 
contributions sensitively depends on the rank of the gauge group, and thus
may contain terms that at first sight would not be allowed in a large $N$ limit of the quiver gauge
theory. The $\mu$ terms, in particular, can be viewed as baryon-type operators for $SU(2)$,
and thus one can easily imagine that they get generated via D-instantons.}
Since the D-instanton contribution decreases exponentially with the volume of the 4-cycles $\Sigma_\alpha$, it would naturally
explain why (some of) the $\mu$-terms are small compared to the string scale.

\bigskip

\bigskip

\newsubsection{Eliminating extra Higgses}

\begin{figure}[t]
\begin{center}
\leavevmode\hbox{\epsfxsize=7.5cm \epsffile{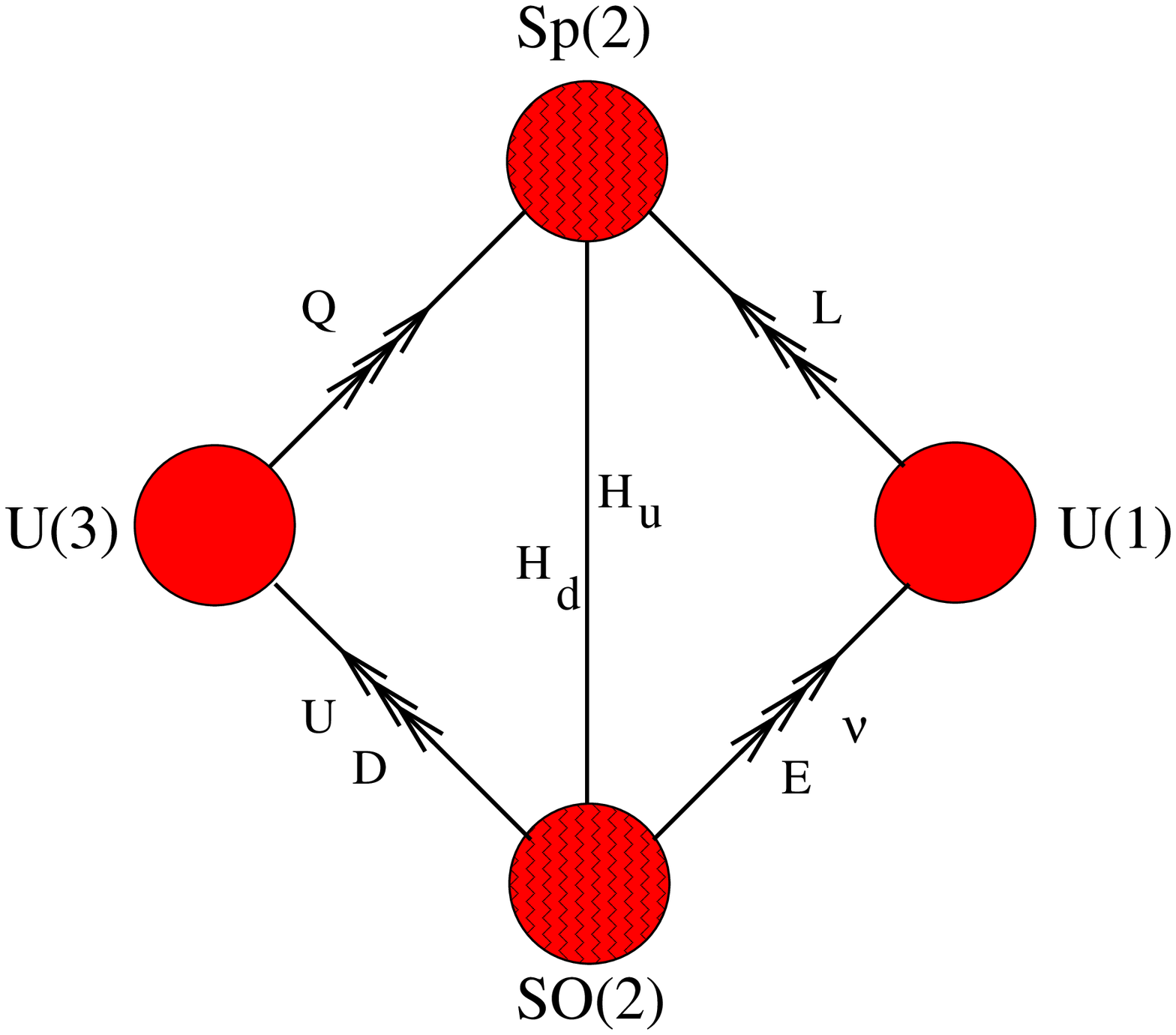}}\\[3mm]
\bigskip
\medskip

\parbox{15.5cm}{Figure 7: An MSSM-like quiver gauge theory, satisfying all rules for world-volume theories in an unoriented string models.}
\end{center}
\end{figure}

From a phenomenological perspective,  the specific model based on the $dP_8$ 
singularity still has several issues that need to be addressed, before it can become fully
realistic. Most immediately noticeable is the multitude of Higgs fields, and the
fact that supersymmetry is unbroken. Supersymmetry breaking effects may get generated
via various  mechanisms: via fluxes, nearby anti-branes, non-perturbative string physics,
etc.  The structure of the  SUSY breaking and $\mu$-terms are strongly 
restricted by phenomenological constraints, such as the suppression of flavor changing 
neutral currents. However, we see no a priori obstruction to the existence of mechanisms
that would sufficiently lift the masses of all extra Higgses and effectively eliminate them 
from the low energy spectrum.

The presence of the extra Higgs fields is dictated via the requirement (on all
D-brane constructions on orientable CY singularities)  that
each node should have an equal number of in- and out-going lines. To eliminate
this feature, it is natural to look for generalizations  
among gauge theories on orientifolds of CY
singularities. Near orientifold planes, D-branes can support real gauge groups
like $SO(2N)$ or $Sp(N)$. With this generalization, one can draw a more minimal
quiver extension of the SM, with fewer Higgs fields. An example of such a
quiver is drawn in fig 7. It should be straightforward to find an orientifolded CY singularity
and fractional brane configuration that would reproduce this quiver.
The  extra $U(1)$ factors in  fig 7 can then be dealt with in a similar way 
as in our $dP_8$  example.

\bigskip

\bigskip

\noindent
{\large \bf Acknowledgments}

It is a pleasure to thank Vijay Balasubramanian, David Berenstein, Michael Douglas,
Shamit Kachru,  Elias Kiritsis, J\'anos Koll\'ar, John McGreevy, and Michael Schulz for helpful
discussion and comments. This work was
supported by the National Science Foundation under grants PHY-0243680
and DMS-0606578 and by an NSF Graduate Research Fellowship (M.B.). D.M.would like to thank the IHES for hospitality and support when part of this work was done. Any opinions,
findings, and conclusions or recommendations expressed in  this material are
those of the authors and do not necessarily reflect the views of the National Science
Foundation.

\appendix

\renewcommand{\newsection}[1]{
\addtocounter{section}{1} \setcounter{equation}{0}
\setcounter{subsection}{0} \addcontentsline{toc}{section}{\protect
\numberline{\Alph{section}}{{ \rm #1}}} \vglue .6cm \pagebreak[3]
\noindent{\large \bf  \thesection. #1}\nopagebreak[4]\par\vskip .3cm}


\begin{thebibliography}{[AHU]}

\bibitem{ALE}
  M.~R.~Douglas and G.~W.~Moore,
   ``D-branes, Quivers, and ALE Instantons,''
 {\tt hep-th/9603167.}

\bibitem{KW}
  I.~R.~Klebanov and E.~Witten,
   ``Superconformal field theory on threebranes at a Calabi-Yau  singularity,''
  %
  Nucl.\ Phys.\ B {\bf 536}, 199 (1998)
  {\tt hep-th/9807080}.


\bibitem{MoPl}
D.~R.~Morrison and M.~R.~Plesser, ``Nonspherical Horizons. 1,"
Adv.\ Theor.\ Math.\ Phys.\  {\bf 3}, 1 (1999)
{\tt hep-th/9810201}.

\bibitem{swamp}
  C.~Vafa,
  ``The string landscape and the swampland,''
  arXiv:hep-th/0509212.
  
\bibitem{MH}
  H.~Verlinde and M.~Wijnholt,
  ``Building the standard model on a D3-brane,''
  arXiv:hep-th/0508089.

\bibitem{martijn}
M.~Wijnholt,
``Large volume perspective on branes at singularities,''
{\tt hep-th/0212021}.


\bibitem{chris}
  C.~P.~Herzog,
  ``Exceptional collections and del Pezzo gauge theories,''
  JHEP {\bf 0404}, 069 (2004), {\tt hep-th/0310262.}

\bibitem{bergman} A. Bergman and C. Herzog, \lq\lq The volume of non-spherical
 horizons and the AdS/CFT correspondence," arXiv:hep-th/0108020.

\bibitem{Aspinwall}
  P.~S.~Aspinwall,
  ``K3 surfaces and string duality,''
  arXiv:hep-th/9611137.

\bibitem{VafaWitten}
  C.~Vafa and E.~Witten,
  Nucl.\ Phys.\ Proc.\ Suppl.\  {\bf 46}, 225 (1996)
  [arXiv:hep-th/9507050].

\bibitem{Grimm}
  T.~W.~Grimm and J.~Louis,
  ``The effective action of N = 1 Calabi-Yau orientifolds,''
  Nucl.\ Phys.\ B {\bf 699}, 387 (2004)
  [arXiv:hep-th/0403067].

\bibitem{louisCY} H. Jockers and J. Louis \lq The effective action
of D7-branes in $N=1$ Calabi-Yau Orientifolds,'
arXiv:hep-th/0409098.

\bibitem{Cheung:1997az}
  Y.~K.~Cheung and Z.~Yin,
  ``Anomalies, branes, and currents,''
  Nucl.\ Phys.\ B {\bf 517}, 69 (1998)
  [arXiv:hep-th/9710206].

\bibitem{minasian}
  M.~Marino, R.~Minasian, G.~W.~Moore and A.~Strominger,
  ``Nonlinear instantons from supersymmetric p-branes,''
  JHEP {\bf 0001}, 005 (2000)
  [arXiv:hep-th/9911206].

\bibitem{Douglas:2000}
  M.~R.~Douglas, B.~Fiol and C.~Romelsberger,
  JHEP {\bf 0509}, 006 (2005)
  [arXiv:hep-th/0002037].
    
\bibitem{kapustin}
  A.~Kapustin and Y.~Li,
  ``Stability conditions for topological D-branes: A worldsheet approach,''
  arXiv:hep-th/0311101.

\bibitem{Intriligator:2005aw}
  K.~Intriligator and N.~Seiberg,
  ``The runaway quiver,''
  JHEP {\bf 0602}, 031 (2006)
  [arXiv:hep-th/0512347].


\bibitem{mass}
  L.~E.~Ibanez, R.~Rabadan and A.~M.~Uranga,
   ``Anomalous U(1)'s in type I and type IIB D = 4, N = 1 string vacua,''
  %
  Nucl.\ Phys.\ B {\bf 542}, 112 (1999) {\tt hep-th/9808139.}
\bibitem{Antoniadis:2002}
  I.~Antoniadis, E.~Kiritsis and J.~Rizos,
  ``Anomalous U(1)s in type I superstring vacua,''
  Nucl.\ Phys.\ B {\bf 637}, 92 (2002)
  [arXiv:hep-th/0204153].
\bibitem{karpov}
B. V. Karpov\ and\ D. Yu. Nogin,
``Three-block Exceptional Collections over del Pezzo Surfaces,''
Izv. Ross. Akad. Nauk Ser. Mat. {\bf 62} (1998), no. 3, 3--38;
translation in Izv. Math. {\bf 62} (1998), no.~3, 429--463,
{\tt alg-geom/9703027}.


\bibitem{delta27}
  D.~Berenstein, V.~Jejjala and R.~G.~Leigh,
  ``The standard model on a D-brane,''
  Phys.\ Rev.\ Lett.\  {\bf 88}, 071602 (2002)
  [arXiv:hep-ph/0105042].

\bibitem{Aldazabal}
  G.~Aldazabal, L.~E.~Ibanez, F.~Quevedo and A.~M.~Uranga,
   ``D-branes at singularities: A bottom-up approach to the string  embedding of
  the standard model,''
  JHEP {\bf 0008}, 002 (2000)
  [arXiv:hep-th/0005067].
\bibitem{martijn2}
  M.~Wijnholt,
  ``Parameter space of quiver gauge theories,''
  arXiv:hep-th/0512122.
\bibitem{pol}
J.~Polchinski, ``Tensors from {K3} orientifolds,'' Phys. Rev. D {\bf 55}
  (1997) 6423--6428, {\tt hep-th/9606165}.

\bibitem{dgm}
M.~R. Douglas, B.~R. Greene, and D.~R. Morrison, ``Orbifold resolution by
  {D}-branes,'' Nucl. Phys. B {\bf 506} (1997) 84--106, {\tt
  hep-th/9704151}.

\bibitem{ddg}
D.~E. Diaconescu, M.~R. Douglas, and J.~Gomis, ``Fractional branes and wrapped
  branes,'' J. High Energy Phys. {\bf 02} (1998) 013, {\tt
  hep-th/9712230}.

\bibitem{lnv}
A.~Lawrence, N.~Nekrasov, and C.~Vafa, ``On conformal field theories in four
  dimensions,'' 
Nucl. Phys. B {\bf 533} (1998) 199--209, {\tt hep-th/9803015}.




\bibitem{eighteen}
  V.~Balasubramanian, P.~Berglund, J.~P.~Conlon and F.~Quevedo,
   ``Systematics of moduli stabilisation in Calabi-Yau flux
  compactifications,''
  JHEP {\bf 0503}, 007 (2005)
  [arXiv:hep-th/0502058].


\bibitem{Diaconescu}
  D.~E.~Diaconescu, B.~Florea, S.~Kachru and P.~Svrcek,
  ``Gauge - mediated supersymmetry breaking in string compactifications,''
  JHEP {\bf 0602}, 020 (2006)
  [arXiv:hep-th/0512170].



\bibitem{kawamata}
Y. Kawamata, ``{Crepant blowing-up of 3-dimensional canonical singularities
and its application to degenerations of surfaces}'', Ann. of Math. (2) {\bf 127}
(1988), 93--163.

\bibitem{mori}
S. Mori,``{Threefolds whose canonical bundles are not numerically 
effective}'', Ann. of Math. (2) {\bf 116} (1982), 133--176.

\bibitem{morrison-seiberg}
D. R. Morrison and N. Seiberg,
``{Extremal transitions and five-dimensional
  supersymmetric field theories}'', Nuclear Phys. B {\bf 483} (1997), 229--247,
  {\tt arXiv:hep-th/9609070}.

\bibitem{greene-morrison-vafa}
B.~R.~Greene, D. R. Morrison, and C.~Vafa, 
``{A geometric realization of confinement}'',
  Nuclear Phys. B {\bf 481} (1996), 513--538, {\tt arXiv:hep-th/9608039}.

\bibitem{Witten:1996bn}
  E.~Witten,
  ``Non-Perturbative Superpotentials In String Theory,''
  Nucl.\ Phys.\ B {\bf 474}, 343 (1996)
  [arXiv:hep-th/9604030].
  
\bibitem{Bershadsky:1996gx}
  M.~Bershadsky, A.~Johansen, T.~Pantev, V.~Sadov and C.~Vafa,
  ``F-theory, geometric engineering and N = 1 dualities,''
  Nucl.\ Phys.\ B {\bf 505}, 153 (1997)
  [arXiv:hep-th/9612052].
\bibitem{Ganor}
  O.~J.~Ganor,
  ``A note on zeroes of superpotentials in F-theory,''
  Nucl.\ Phys.\ B {\bf 499}, 55 (1997)
  [arXiv:hep-th/9612077].
 
  
\bibitem{Beasley:2004ys}
  C.~Beasley and E.~Witten,
  ``New instanton effects in supersymmetric QCD,''
  JHEP {\bf 0501}, 056 (2005)
  [arXiv:hep-th/0409149].

\bibitem{Blumenhagen:2006}
  R.~Blumenhagen, M.~Cvetic and T.~Weigand,
   ``Spacetime instanton corrections in 4D string vacua - the seesaw mechanism
  arXiv:hep-th/0609191.

\bibitem{Ibanez:2006}
  L.~E.~Ibanez and A.~M.~Uranga,
  arXiv:hep-th/0609213.

\bibitem{FKMS}
  B.~Florea, S.~Kachru, J.~McGreevy, and N. Saulina, 
  ``Stringy instantons and quiver gauge theories'' [arXiv:hep-th/0610003].

\bibitem{Berenstein:2006}
  D.~Berenstein,
  ``Branes vs. GUTS: Challenges for string inspired phenomenology,''
  arXiv:hep-th/0603103.
\end{thebibliography}
\end{document}